\documentclass[english,aps, prl, superscriptaddress, citeautoscript, reprint, twocolumn, 10pt, tightenlines]{revtex4-1}
\usepackage{amsmath, amssymb, graphicx, bm}
\usepackage[colorlinks=true,citecolor=blue,urlcolor=blue]{hyperref}
\usepackage{xcolor, soul}
\usepackage{gensymb}
\usepackage{bbm}
\usepackage{xr}
\usepackage{float}
\usepackage{bbm}

\makeatletter
\newcommand*{\addFileDependency}[1]{
  \typeout{(#1)}
  \@addtofilelist{#1}
  \IfFileExists{#1}{}{\typeout{No file #1.}}
}
\makeatother

\newcommand*{\myexternaldocument}[1]{
    \externaldocument{#1}
    \addFileDependency{#1.tex}
    \addFileDependency{#1.aux}
}

\myexternaldocument{SuppInfo}

\usepackage{scalerel}

\def\be{\begin{equation}}
\def\ee{\end{equation}}
\def\bmu{\begin{multline}}
\def\bea{\begin{eqnarray}}
\def\eea{\end{eqnarray}}
\def\nn{\nonumber}

\def\n{{\bf n}}
\def\v{{\bf v}}
\def\p{{\bf p}}

\def\R{{\bf R}}
\def\br{{\bf r}}
\def\pa{\partial} 
\def\pt{\partial_t} 

\def\f{\frac}
\def\a{\alpha}

\def\l{\left(}
\def\r{\right)}
\def\[{\left[}
\def\]{\right]}

\newcommand{\me}{\mathrm{e}}

\begin{document}

\title{Active ploughing through a compressible viscoelastic fluid: Unjamming and emergent nonreciprocity}
\author{Jyoti Prasad Banerjee}
\affiliation{Simons Centre for the Study of Living Machines, National Centre for Biological Sciences (TIFR), Bangalore, India}
\author{Rituparno Mandal}
\affiliation{Institute for Theoretical Physics, Georg-August-Universit\"at G\"ottingen, 37077 G\"ottingen, Germany }
\author{Deb Sankar Banerjee}
\affiliation{Department of Physics, Carnegie Mellon University, Pittsburgh, USA}
\author{Shashi Thutupalli} 
\affiliation{Simons Centre for the Study of Living Machines, National Centre for Biological Sciences (TIFR), Bangalore, India}
\affiliation{International Centre for Theoretical Sciences (TIFR), Bangalore, India}
\author{Madan Rao}
\affiliation{Simons Centre for the Study of Living Machines, National Centre for Biological Sciences (TIFR), Bangalore, India}

\date{\today}

\begin{abstract}
A dilute suspension of active Brownian particles in a dense compressible viscoelastic fluid, forms a natural setting to study the emergence of nonreciprocity during a dynamical phase transition. At these densities, the transport of active particles is strongly influenced by the passive medium and shows a dynamical jamming transition as a function of activity and medium density. 
 In the process, the compressible medium is actively churned up -- for low activity, the active particle gets self-trapped in a spherical cavity of its own making, while for large activity, the active particle ploughs through the medium, either accompanied by a moving anisotropic wake, or leaving a porous trail. A hydrodynamic approach makes it evident that the active particle generates a long range density wake which breaks fore-aft symmetry, consistent with the simulations. Accounting for the back reaction of the compressible medium leads to (i) dynamical  jamming of the active particle, and (ii) a dynamical {\it non-reciprocal} attraction between two active particles moving along the same direction, with the trailing particle catching up with the leading one in finite time. We emphasize that these nonreciprocal effects appear only when the active particles are moving and so manifest in the vicinity of the jamming-unjamming transition.
\end{abstract}

\maketitle

\section{Introduction}
\label{sec:intro}

There is a lot of interest in the effective long-range interactions that emerge amongst active particles moving through a dynamically responsive medium. Examples include motile particles embedded in a Stokesian fluid~\cite{Vishen2019} or on an elastic substrate~\cite{SRAS2020} and diffusiophoretic flows in a viscous suspension of chemically active particles~\cite{SSSR2014}. The absence of time reversal symmetry manifests in large density fluctuations at steady state~\cite{ASSRJT, RMP2013}, and  in the appearance of nonreciprocal interactions between particles~\cite{JDMRSR2002,JDMRSR2004,KHMR,SSSR2014,Saha2020,SRAS2020,You2020,vitelli2021}.

In this paper we ask whether nonreciprocal interactions could emerge as a result of a dynamical phase transition. To realise such 
emergent nonreciprocity, we 
study a dilute collection of motile active particles embedded in a dense compressible fluid suspension close to dynamical arrest. 
Our results are based on (a) numerical simulations of an agent-based model, and (b) analytical and numerical treatments of hydrodynamic equations. 
We find that there is a dynamical feedback between the motility of the active particles and the corresponding slow remodelling of the passive compressible medium. 
Such active particles are {\it ploughers}, as opposed to {\it cruisers} whose motility speed is unaffected by the medium, e.g.~\cite{SRAS2020}.
As a result, ploughers exhibit a jamming-unjamming transition at fixed medium density. We find that in the unjammed phase, the moving active particles develop a dynamic nonreciprocal interaction with each other arising from the compressibility of the passive medium. 
We emphasize that this long-range nonreciprocal sensing appears only when the active particles are moving through a momentum non-conserving medium, and consequently shows up in the neighbourhood of the jamming-unjamming transition.

Our study reveals a hitherto unappreciated facet in this intensely researched field of dense active assemblies~\cite{Mandal2016, RMandal2020, Henkes, Ni2013, BerthierSzamel, BerthierKurchan2013, Berthier2014}, where the focus has been on fluidisation~\cite{Mandal2016}, intermittency~\cite{RMandal2020} and jamming~\cite{RMandal2020,Henkes} close to glass transition. The current work should be relevant to a variety of cellular and non-cellular contexts, where the medium is dense but compressible, pliable but slow to relax. Such situations can occur in the
(i) transport of constituent or embedded particles in the cytoplasm~\cite{Parry, Nishizawa}, (ii) facilitated transport of transcription factories and exogenous particles 
embedded within the cell nucleus~\cite{Feroz}, (iii) movement of bacteria and cancer cells in fabricated soft porous media or in tissues~\cite{Tapomoy, Tapomoy2, Weitz, Shao, Thirumalai}, (iv) burrowing movement of ants and worms in dense soil~\cite{Ant, Ant2, Ant3, Earthworm}, and (v) intrusion of active particles in a disordered bubble raft or a dense suspension of solid colloidal particles~\cite{Bechinger, SinghMangal}.

\section{Dynamics of active Brownian particles in a passive medium}

\begin{figure*}[t!]
\centering
\includegraphics[width=0.9\textwidth]{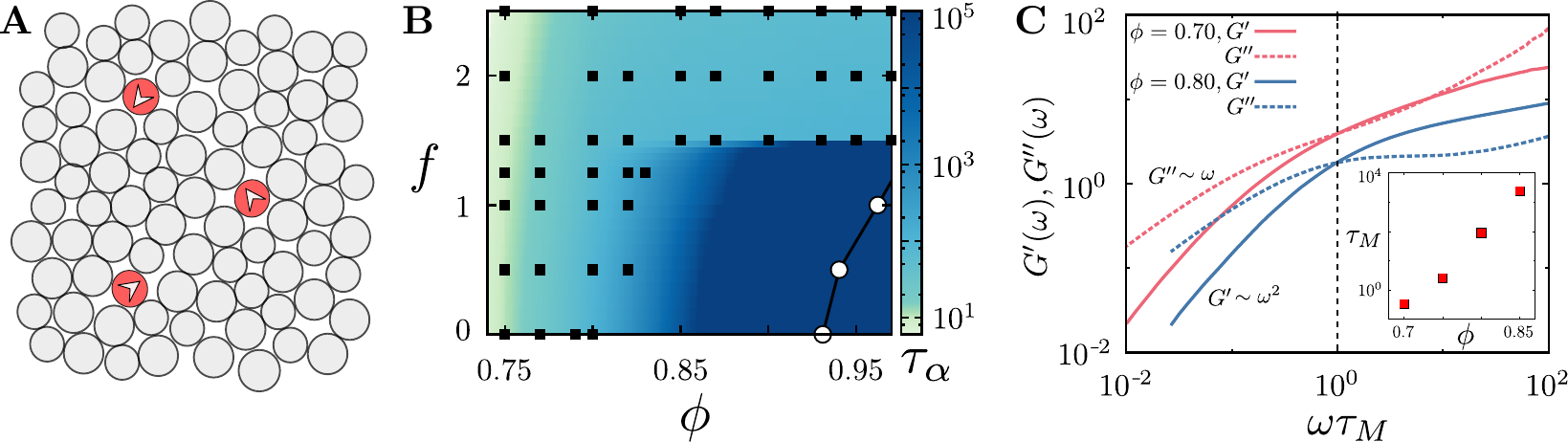}
\caption{\textbf{Active motile particles in a dense medium -- approach to glass and viscoelasticity.} \textbf{A} Schematic of dilute suspension of self-propelled particles of area fraction $\phi_a$ (red particles with arrows showing instantaneous direction  ${\bf n}$ of propulsive force $f ({\bf n}$) moving through a dense compressible passive fluid of area fraction $\phi$ (grey particles). \textbf{B} 
Dynamical phase diagram in the $f-\phi$ plane for fixed $T= 0.5$, $\phi_a=0.017$ and $\tau= 50$, showing the macroscopic liquid and solid (glass) phases adjoining the cage-hopping ``super-cooled liquid'' regime, as determined from the $\a$-relaxation time, $\tau_\a$  ({\it Supplementary Information} Sec.~S2). The glass transition at density $\phi_{\scaleto{VFT}{2.5pt}}(f)$ (open circles) is obtained by fitting $\tau_\a$ to a Vogel-Fulcher form ({\it Supplementary Information} Sec.~S2). \textbf{C} Frequency dependence of the elastic 
 $G'$ and viscous $G''$ responses, at different values of $\phi$, shows that the passive medium behaves as a Maxwell viscoelastic fluid with relaxation time $\tau_{\scaleto{M}{3pt}}$. (inset) $\tau_{\scaleto{M}{3pt}}$ as a function of area fraction $\phi$ increases exponentially close to the glass transition.}
  \label{fig:fig1}
\end{figure*}

Our model of the 2d background passive medium is similar to the Kob-Andersen~\cite{Kob, Bruning2008} binary mixture of soft spheres with volume fraction $\phi$ so as to be able to tune it across a glass transition
at constant temperature $T$. To this passive medium, we add a dilute amount $\phi_a \ll \phi$ of active Brownian soft particles (ABPs) ~\cite{Fily2012,Takatori2015,Codina2017,Cates2015} which are made motile 
by assigning to them independent
random forces ${\bf f} = f\, \mathbf{n}$, whose orientation $\mathbf{n}\equiv(\cos \theta, \sin \theta)$ is exponentially correlated over a  persistence time
$\tau$. The dynamics of all the interacting particles labelled $i$ are described by a 
Langevin equation subject to a thermal noise $\vartheta$ of zero mean and variance equal to $2 \gamma k_B T$, while the subset $i \in \mathcal{A}$ of ABPs are subject to
additional active stochastic forces,
\bea
m{\ddot{\mathbf{x}}}_i & = & -\gamma {\dot{\mathbf{x}}}_i - \sum_{i\neq j}^{N} \pa_j V_{ij} + f \mathbf{n}_i \,\mathbbm{1}_{(i\in\mathcal{A})}+ \mbox{\boldmath $\vartheta$}_i\,, \nn \\
\dot{\theta}_i & = & \xi_i\,\,\,\,\,\,\,\,\, \mbox{for} \,\,{i\in\mathcal{A}}\,.
\label{eq:abp}
\eea

where $\mathbbm{1}_{(i\in\mathcal{A})}$ is the indicator function which ensures that the active forces are restricted to particles $i$ in the active set $\mathcal{A}$.
The orientation angle ${\theta}_i$ undergoes rotational diffusion described by an
athermal noise $\xi_i$, with  zero mean and correlation $\langle \xi_i(t) \xi_j(t') \rangle = 2 \tau^{-1} \delta_{ij} \delta(t-t')$. 
Its effect on the ${\mathbf{x}}_i$-dynamics is an exponentially correlated vectorial noise with correlation time $\tau$,
which being unrelated to the drag $\gamma$, violates the fluctuation-dissipation relation.
The inter-particle potential $V_{ij}$ is taken to be a purely repulsive (inverse power law (IPL)) with particle diameter $\sigma$ (details of the simulation appear in {\it Supplementary Information} Sec.~S1). We work in the high friction limit, where particle inertia can be ignored.

Starting from homogeneous and isotropic initial conditions, 
we evolve the system to its steady state by integrating over a time of order $10^2$\,$\tau_{\a}$, where $\tau_{\a}$ is the density relaxation time, also called $\a$-relaxation time ({\it Supplementary Information} Sec.~S2). Throughout, we work in the low temperature regime
$T = 0.5, 10^{-1}, 10^{-3}$, over a wide range of densities 
$\phi \in \[0.08, 0.97\]$, and scan through a broad range of the active parameters $f$ and $\tau$.

\section{Interplay between active self-propulsion and viscoelasticity of the medium}

\begin{figure*}[t!]
\centering
\includegraphics[width=0.8\textwidth]{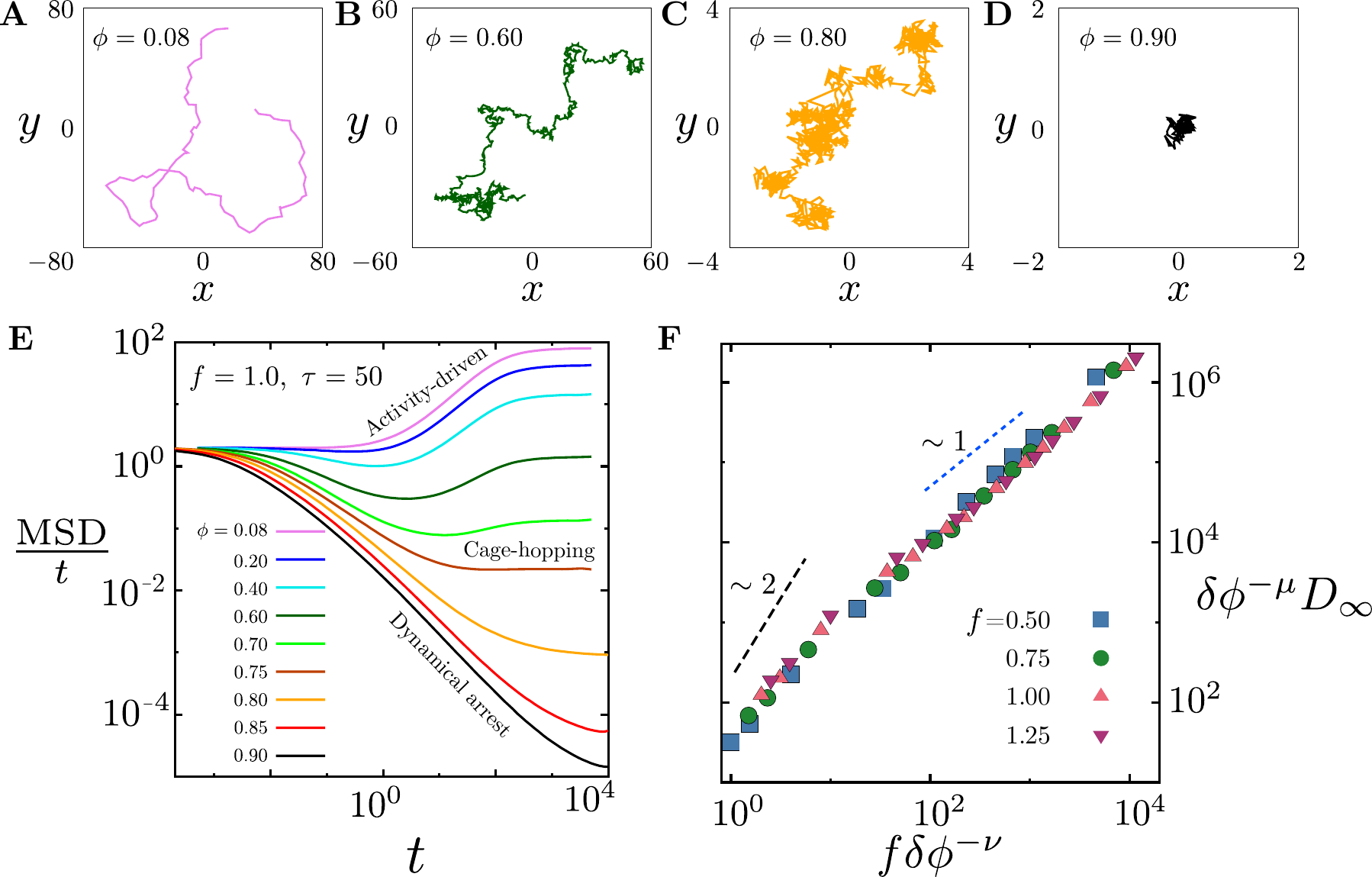}
\caption{ \textbf{Crossover in transport characteristics of minority active particles.}
\textbf{A-D} Typical trajectories of the active particle as a function of $\phi$ at fixed $f=1$ and $\tau=50$, showing (i) activity-dominated transport,
(ii) glass dominated cage-hopping transport and (iii) dynamical arrest, recorded over a time $t = 500$.
\textbf{E} Mean square displacement (scaled by time) computed from active particle trajectories suggests a crossover as a function of $\phi$. \textbf{F} Crossover scaling collapse of the late time diffusion coefficient $D_{\infty}(\phi, f)$, described in Eq.\,\ref{eq:crossover}, in the scaling variable $y\equiv f/\delta \phi^{\nu}$, 
where $\delta \phi$ ($0.026 \leq \delta \phi \leq 0.770$) is the deviation from the MCT
value, $\phi_{\scaleto{MCT}{2.5pt}}(f)$. The scaling exponents are found to be $\mu\approx 6.5$ and
$\nu\approx 2.5$.  The dashed lines suggest a crossover of the scaling function from 
${\cal D}(y) \to y^2$ to ${\cal D}(y) \to y$, as $y$ increases.} 
  \label{fig:fig2}
\end{figure*}

Fig.\,\ref{fig:fig1}\textbf{A} shows a schematic of a dilute suspension ($\phi_a \ll \phi$) of self-propelled particles moving through a dense compressible medium. While the macroscopic structural 
properties of such dense assemblies are rather innocuous,  their dynamical features display characteristic slow relaxation, aging~\cite{Mandal2020} and dynamical arrest~\cite{Mandal2016} as the density $\phi$ is increased.  The dynamics of the medium at large space and time scales, is summarized in a phase diagram (Fig.\,\ref{fig:fig1}\textbf{B}) in the $f-\phi$ plane (for fixed $T$, $\phi_a$ and $\tau$). The phase diagram is
constructed by computing the $\a$-relaxation time $\tau_\a$ from the decay of the density overlap function $Q(t)$
({\it Supplementary Information} Sec.~S2) using the definition $Q(\tau_\alpha)=1/e$. This phase diagram clearly shows the macroscopic liquid and solid (glass) phases adjoining
the cage-hopping ``super-cooled'' regime; fitting  $\tau_\a$ to a Vogel-Fulcher form ({\it Supplementary Information} Sec.~S2) provides an estimate for the glass transition density $\phi_{\scaleto{VFT}{2.5pt}}(f)$ (Fig.\,\ref{fig:fig1}\textbf{B})~\cite{Mandal2016}.

Typical of an approach to a glass, the  mean square displacement (MSD) averaged over all the passive particles shows a plateauing and cage hopping dynamics, as the density $\phi$ is increased (Fig.\,S1)~\cite{Bechinger, BerthierBiroli, Zaccarelli, ZaccarelliPoon}. From these graphs we extract the long time diffusivity $D_{\infty}$ ({\it Supplementary Information} Sec.~S2). In the limit $\phi_a \ll \phi$ and $\tau$ small, we may deduce the linear microrheological properties of the passive medium from the Fourier transform of the MSD~\cite{Mason}, with an effective temperature obtained from the mean kinetic energy of the passive particles. Fig.\,\ref{fig:fig1}\textbf{C} clearly shows that the medium is a viscoelastic Maxwell fluid, with the elastic response $G^{\prime}\sim \omega^2$ and the viscous response $G^{\prime \prime}\sim \omega$, for small $\omega$, where the crossover timescale $\tau_{\scaleto{M}{3pt}}$ increases exponentially with the increase in area fraction $\phi$ close to the glass transition.

We now turn our attention to the minority component, the small fraction of motile active particles --  Fig.\,\ref{fig:fig2}\textbf{A-D} shows typical trajectories of the active motile particles at increasing values of $\phi$, keeping
$f$ and $\tau$ fixed. The density of the passive medium affects the transport of the active particles -- thus at low density $\phi$, the motile particles 
show an {\it activity-dominated} transport (Fig.\,\ref{fig:fig2}\textbf{A},\textbf{B}), which crosses over to a {\it cage-hopping dominated} transport (Fig.\,\ref{fig:fig2}\textbf{C}), as $\phi$ increases. As $\phi$ increases further, while still being less than $\phi_{\scaleto{VFT}{2.5pt}}(f)$, the active particles get dynamically arrested, (Fig.\,\ref{fig:fig2}\textbf{D}). The plot of the MSD of the active particles  for the different values of $\phi$ (Fig.\,\ref{fig:fig2}\textbf{E}), suggests a crossover collapse from activity dominated diffusion proportional to $f^2 \tau$ to a glass dominated cage hopping diffusion with a Vogel-Fulcher form to finally, dynamical arrest. We verify this using a crossover scaling form for the late time diffusion coefficient (Fig.\,\ref{fig:fig2}\textbf{F})
\be
D_{\infty}(\phi, f) = \delta \phi^{\mu} \,{\cal D}\l
 y\equiv \f{f}{\delta \phi^{\nu}}\r
 \label{eq:crossover}
\ee
with $\delta \phi = \phi_{\scaleto{MCT}{2.5pt}}(f) - \phi$, the deviation from the mode-coupling transition. The excellent collapse
with exponents $\mu\approx 6.5$ and $\nu \approx 2.5$, suggests `critical behaviour' at the mode-coupling transition, $\phi_{\scaleto{MCT}{2.5pt}}(f) < \phi_{\scaleto{VFT}{2.5pt}}(f)$. The asymptotic behaviour of the crossover scaling function ${\cal D}(y)$ at small $y$ (Fig.\,\ref{fig:fig2}\textbf{F}), suggests that $D_{\infty} \approx f^2 \, \delta \phi^{3/2}$.

We remark on the connection between the crossover scaling of the MSD of active particles as a function of density of the passive medium with recent observations on crossover behaviour of bacterial motility in a three dimensional porous medium as a function of porosity~\cite{Tapomoy}. In as much as our study applies to this bacterial motility context, we suggest that the reported crossover in~\cite{Tapomoy} reflects a phenotypic change arising from a coupling of the normal bacterial movement to the physical properties of the dense passive medium. 

\section{Remodelling of the compressible viscoelastic medium by the motile particles}

\begin{figure*}[t!]
\centering
\includegraphics[width=0.8\textwidth]{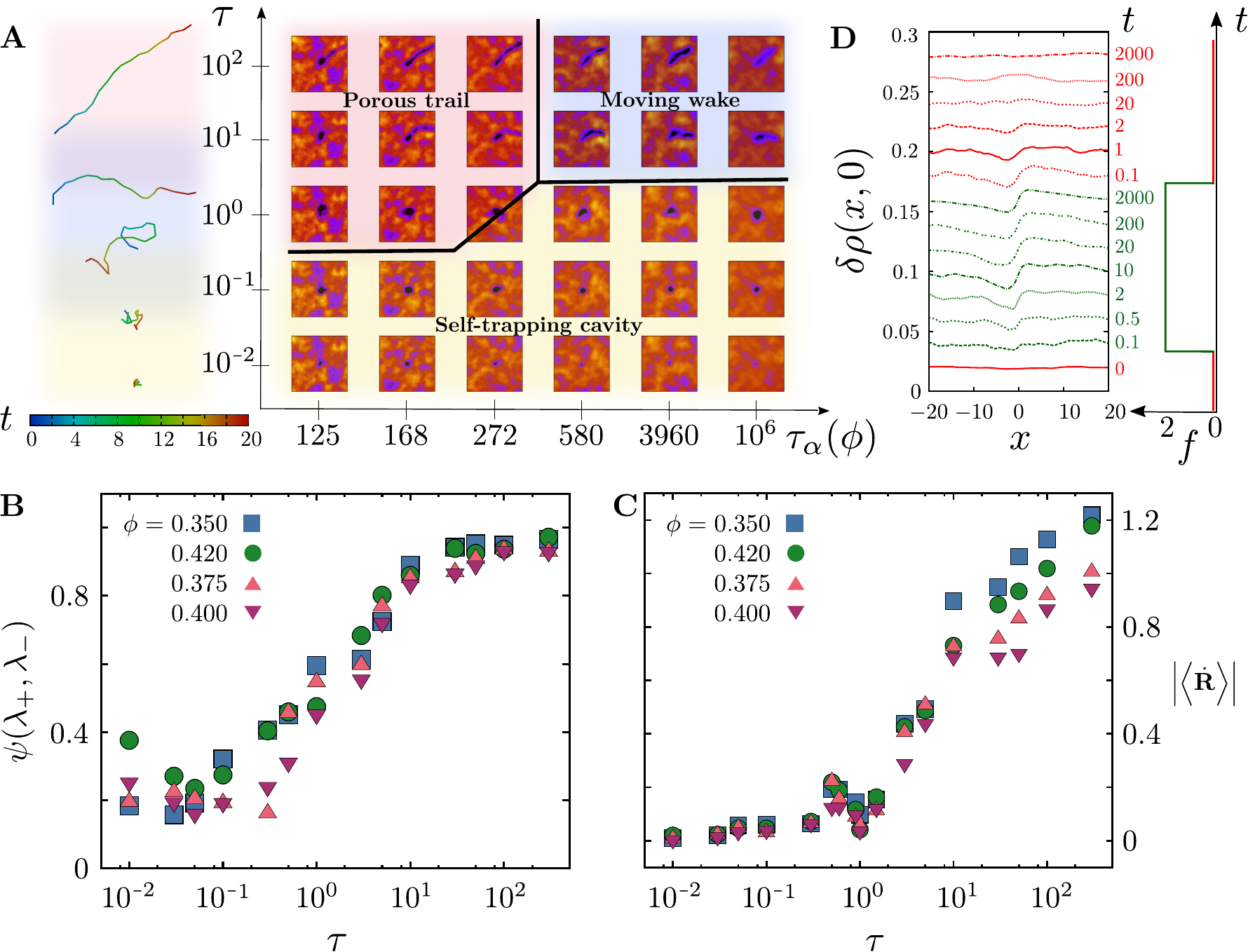}
\caption{\textbf{Remodelling of the compressible medium by the active particles.} \textbf{A} Underdense regions in the compressible medium (darker blue colours indicate low $\phi$, and the yellow colours indicate a high $\phi$) that is remodelled by the motile particles, as a function of the persistence time $\tau$ and density relaxation time $\tau_\a(\phi)$, for fixed (large) $f$. The geometry of the underdense regions goes from being a static cavity to a moving wake around the motile particle, to a long-lived porous trail. The associated trajectories of the active particles at fixed $\tau_\a(\phi)$ and varying $\tau$ are displayed on the left. The background colours outlines the particle trajectories corresponding to the respective regions marked on the phase diagram. \textbf{B} The geometry of the underdense regions is characterised by a {\it shape parameter}, $\psi = \frac{\lambda_{+} - \lambda_{-}}{\lambda_{+} + \lambda_{-}}$, where $\lambda_{\pm}$ are the eigenvalues of the moment of gyration tensor ({\it{Supplementary Information}} Sec.~S3), and goes from being circular ($\psi \approx 0$) to elongated ($\psi \approx 1$) as $\tau$  increases. \textbf{C} Mean velocity of the active particle (averaged over a time interval $\Delta t = 20$), as a function of the persistence time $\tau$ shows a dynamical transition at $\tau \approx 1$, below which it gets self-trapped in a cavity of its own making.
\textbf{D} Dynamical response of the passive medium recorded at different time points, to a step active force from a single active particle (shown at right), measured in the frame of reference of the moving particle. The response is fore-aft asymmetric and relaxes slowly on switching off the active force. 
}
 \label{fig:fig3}
\end{figure*}

We see that there is a strong feedback between the nature of active particle transport and the dynamical remodelling  of the  passive medium by the active particles ~\cite{Brenner2017}. This is especially prominent in the 
``super-cooled'' liquid regime above the glass transition, where the
active motile particles churn up the medium, inducing large density 
fluctuations that result in long lived density modulations that back-react on the transport of the active particles. For a fixed active force $f$ and temperature $T$, the physical characteristics of the under-dense regions are a result of the interplay between the active driving time $\tau$ and the $\phi$-dependent density relaxation time, $\tau_\a(\phi)$.

Associated with a typical trajectory of the active particles shown in 
Fig.\,\ref{fig:fig3}\textbf{A}, we generate a density map of the medium in the vicinity of the active particle, as a function of $\tau$ and $\tau_\a(\phi)$, keeping $f$ large ($f=3.0$) and $T$ low $(T=10^{-3})$. The geometry and dynamics
of the under-dense regions created by the active particles, shows striking  variations -- (i) a halo (density wake) that 
moves with the motile particle, (ii) a static cavity that traps the active particle and 
(iii) a long-lived porous and tortuous trail as the active particle ploughs through the medium. In Fig.\,\ref{fig:fig3}\textbf{B}, we show how the shape of the under-dense region sharply changes from circular to elongated as a 
function of $\tau$. This geometrical transition appears to coincide with a dynamical transition in the active particle transport -- Fig.\,\ref{fig:fig3}\textbf{C} shows that the speed of the active particle $| \langle \dot{ {\bf R}}\rangle |$ goes from being
non-zero (where the active particle ploughs through the medium)
to zero (where the active particle is self-trapped in a quasi-circular cavity of its own making), as $\tau$ decreases. 

The profile and lifetime of the under-dense regions upon active remodelling,
is a dynamical imprint of the transiting
active particle (both its magnitude and direction) on the medium. 
In Fig.\,\ref{fig:fig3}\textbf{D}, we activate only one of the particles of the medium, by imposing a
step active force for a fixed duration. We measure the dynamical response
of the passive medium (the change in local density from its initial uniform profile) from the start of the activity, {\it in the frame of reference of the moving active particle}. We see that the density response $\delta \rho(x,t)$ is fore-aft asymmetric and that this asymmetric profile relaxes slowly on switching off the active force. This demonstrates  that the passive medium (i) is a compressible fluid, and (ii) retains a memory of the moving event (its magnitude and direction) for some time.

In summary, we find that the active particles remodel the passive compressible medium and that the remodelled compressible medium reacts back on the active particle affecting its large scale movement. The back-reaction from the passive medium, can either facilitate movement of the active particle (in the `moving wake' and `porous' regimes) or trap the active particle (in the `active self-trapping' regime). A striking example of such facilitated transport of active particles in a compressible gel is the the ATP-dependent movement of transcription factories that move through the dense nuclear medium of cells~\cite{Feroz}.

\section{Hydrodynamics of active particles moving in a compressible viscoelastic fluid}

\begin{figure*}[t!]
\centering
\includegraphics[width=0.8\textwidth]{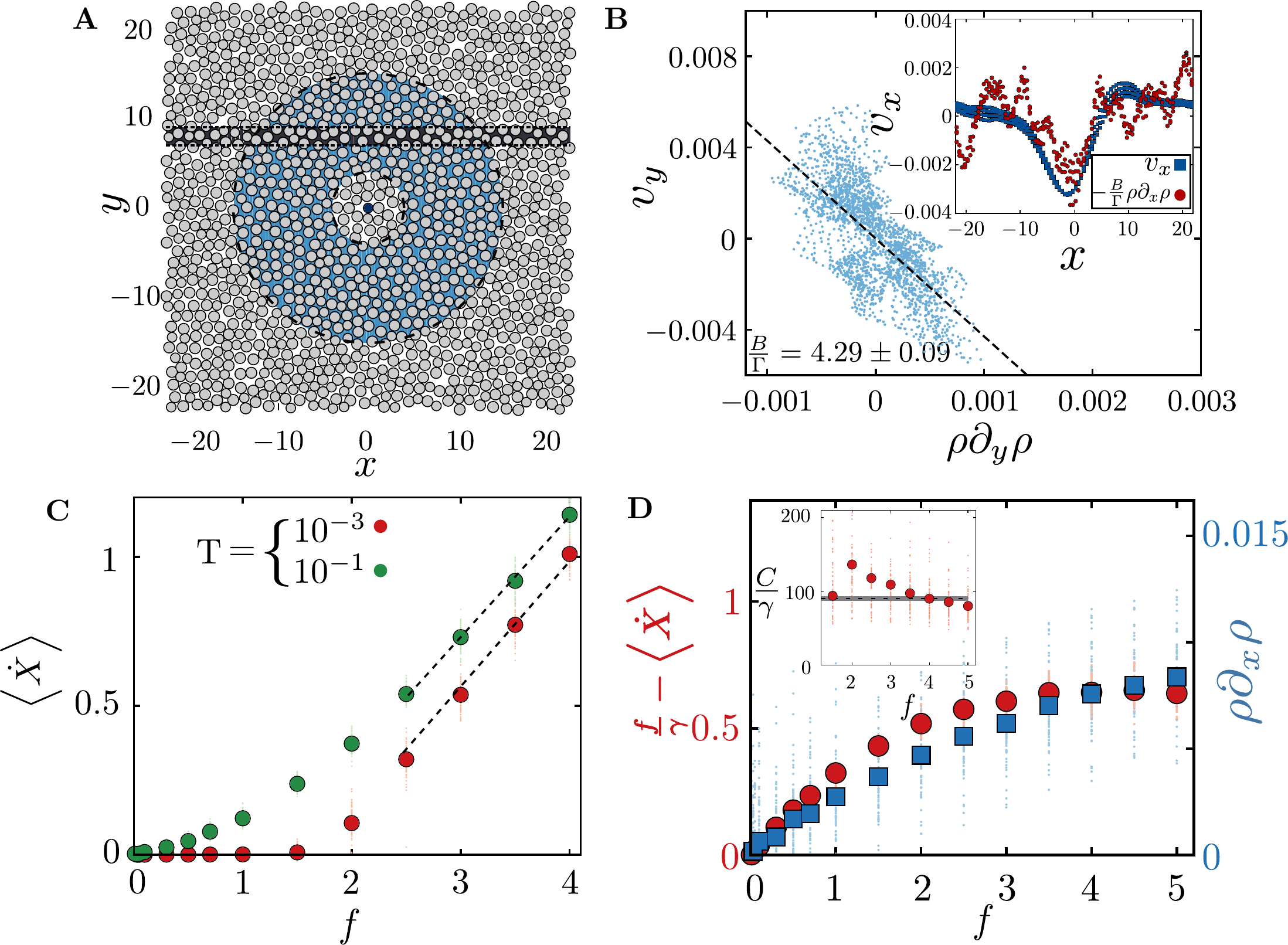}
\caption{\textbf{Verifying hydrodynamic equations by coarse-graining agent-based simulations.}
\textbf{A} Simulation snapshot of the x-y configurations showing an active particle at the origin (black dot) surrounded by particles comprising the compressible medium. We compute the coarse-grained fields, $\rho(\br,t)$,  ${\bf v}(\br,t)$ and their spatial derivatives in the shaded (blue) annular region and the thin shaded (black) rectangular region ({\it Supplementary Information} Sec.~S3). \textbf{B} To verify Eq.\,\ref{eq:forcebalance1}, we plot $v_y(\br,t)$ vs. $\rho \pa_y \rho$ in the shaded region in {\bf A} at a given time; the best-fit line (dashed line) gives the parameter $B/\Gamma = 4.29\pm 0.093$. (inset) Using this value of $B/\Gamma$, we compute 
$v_x(\br,t)$ and  $\rho \pa_x \rho$ versus $x$ along the thin black rectangular region in {\bf A} and find good agreement.
\textbf{C} Mean velocity of the active particle $\dot{X}$ (red dots) vs. active force, $f$, showing the dynamical transition between self-trapping at low $f$ and movement. The mean is obtained by averaging over different initial realisations of the passive medium (with standard deviation displayed). Following Eq.\,\ref{eq:activepart1}, the value of $f/\gamma$ can be extracted from the slope at large $f$.
\textbf{D} To verify Eq.\,\ref{eq:activepart1}, we plot $f/\gamma-\dot{X}$
and $\rho \pa_x \rho$ (just in front of the active particle) versus $f$. This
allows us to compute the parameter $C/\gamma$ (inset), a measure of the back reaction of the medium on the motile particle. The value of $C/\gamma \approx 150$
is an order of magnitude larger than $B/\Gamma$
and drops down to $100$ at larger $f$. 
 }
\label{fig:fig4}
\end{figure*}

For a deeper understanding of the interplay between the movement of the active particles and the asymmetric dynamical response of the compressible passive medium, we
construct a set of active hydrodynamic equations~\cite{RMP2013} and analyse their  solutions in simple situations.

The passive compressible fluid is described by the local density $\rho$ and velocity $\v$ fields, while the dilute collection of active particles $i$ 
with position $\R_i(t)$ are propelled by an active force of magnitude $f$ along their orientations $\n_i(t)$. The density of the medium obeys a
continuity equation,
\be
\pt \rho + \nabla \cdot \l \rho \v \r = 0
\label{eq:density}
\ee
Since the dynamics is overdamped, the local velocity of the medium is obtained by local force balance,

\be
\Gamma \v = \eta \nabla^2 \v - B \rho \nabla \rho + \sum_{i\in\mathcal{A}} f \,\n_i(t) \delta\l {\bf r}- \R_i(t)\r 
\label{eq:forcebalance1}
\ee 
where the velocity of the compressible fluid is driven by 
the body-forces $f\,\n_{i\in\mathcal{A}}$ imposed by the moving active particles. Note that this contribution is present, since the dynamics takes place in a medium that does not conserve momentum.

The second term on the right represents the forces that oppose the 
movement of the passive particles that are being pushed by the active body-forces. These come from an active pressure, which to leading order
arises from a local compressibility of the passive fluid, $P \propto \rho^2
+ \ldots$, due to inter-particle interactions
(at low $T$, the linear contribution is insignificant). The ``compressibility'' $B$,
with units of {\it force} $\times$ {\it length}, is positive
and can in principle depend on $\rho$.

The other terms correspond to the usual momentum dissipation, a viscous contribution $\eta$ coming from 
collisions with the passive particles and friction $\Gamma$ arising from both collisions with other particles and from the ambient medium.
Importantly, in the high density regime approaching the glass transition, these kinetic coefficients $\eta$ and $\Gamma$ may be 
strongly dependent on the local density $\rho$.

The dynamics of the active particles in the overdamped limit is also given by
force balance. In the dilute limit, low $\phi_a$, when there are no direct interactions between active particles, the balance is again 
between the propulsive body-forces and the local pressure due to the compressible fluid,
\bea
\label{eq:activepart1}
\gamma \dot \R_i & = & \,f \n_i - C \rho \nabla \rho \Big \vert_{at\,\, i} \\
\dot \n_i & = & \mbox{{\boldmath $\xi$}}_i(t)
\label{eq:activepart2}
\eea
for all ${i\in\mathcal{A}}$ and the vectorial noise {\boldmath $\xi$}$_i$ has zero mean and is exponentially correlated in time, with a persistence time $\tau$. Equation\,\ref{eq:activepart1} accounts for the back-reaction of the medium on the dynamics of the active particles, which feels a block coming from particle pile up ahead of it. We have assumed in Eq.\,\ref{eq:activepart2} that the direction of the propulsion force is set by some internal detailed-balance violating mechanism,  
independent of the passive medium.
Note that the active compressibility $C$ is positive, dependent on $\rho$, and could be different from $B$. Further, in the limit of low $\phi_a$, one expects $\gamma$ to be a single particle friction, while $\Gamma$ to be a collective frictional dissipation; the latter could be high when $\rho$ is large. 

We will refer to such polar active particles as {\it ploughers}, as opposed to {\it cruisers}, whose speed is unaffected by the medium, e.g.~\cite{SRAS2020}.

Note that although the passive and active particles have comparable sizes, we 
are treating the passive medium using a coarse-grained density, but the active particles as ``point-particulate''. Thus our hydrodynamic description should be valid over scales larger than a few particle sizes. 


We now check whether the continuum hydrodynamic equations, Eqs.\,\ref{eq:density}-\ref{eq:activepart1}, describe in a coarse-grained sense, the agent-based dynamics represented by Eq.\,\ref{eq:abp}.
For this we compute the local coarse grained density and velocity fields of the passive fluid from our simulation trajectories, using an interpolation and smoothing scheme ({\it Supplementary Information} Sec.~S3). 

Take the case of a single particle, with no orientational fluctuations. The results appear
in Fig.\,\ref{fig:fig4}\textbf{A-C}: the relation between the local velocity of the medium and the pile up of the density embodied in Eq.\,\ref{eq:forcebalance1} is shown to hold up in Fig.\,\ref{fig:fig4}\textbf{B}, even making allowance for a possible density dependent 
coefficient $B/\Gamma$ (the contribution from viscous dissipation is significantly lower than the rest and so we drop it).  Similarly the form of the back reaction embodied in Eq.\,\ref{eq:activepart1} is
also borne out in Fig.\,\ref{fig:fig4}\textbf{C,D}, albeit with a density (or force) dependent $C/\gamma$ (inset Fig.~\ref{fig:fig4}\textbf{D}).
Note that consistent with our discussion above, 
$C/\gamma$ is an order of magnitude larger than 
$B/\Gamma$. The fact that $C/\gamma$ drops suddenly beyond $f\approx 2.5$, would suggest that the friction experienced by the active particle increases with increasing $f$ and then saturates to a constant value. In principle, if this drop is large enough, this could lead to an active discontinuous shear thickening~\cite{WyartCates}.

\subsection{Linearised hydrodynamics of a single active particle moving in a compressible medium}

\begin{figure*}[t!]
\centering
\includegraphics[width=0.95\textwidth]{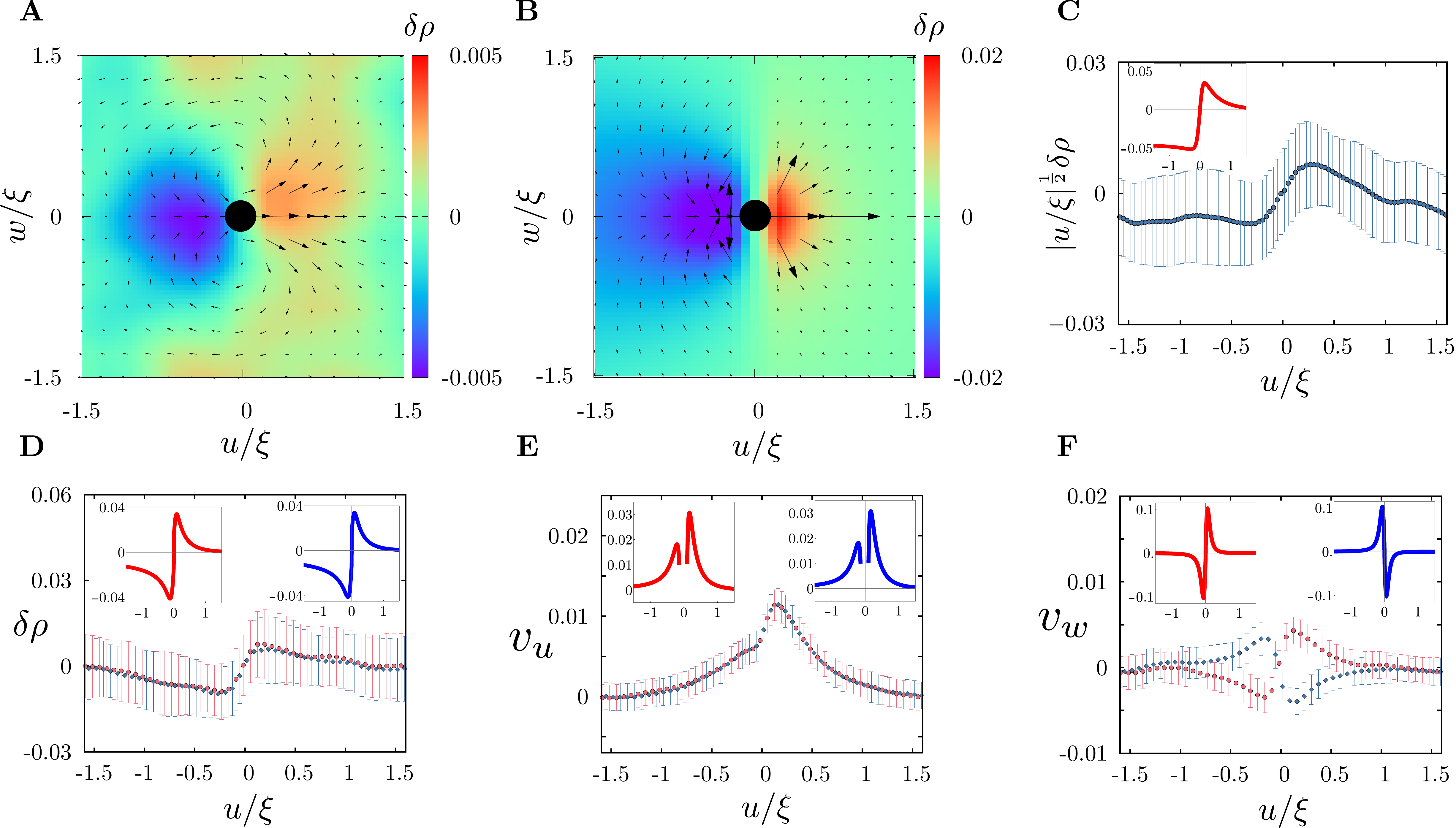}
\caption{\textbf{Density profile and flow field surrounding a single active particle moving through the passive medium.}
\textbf{A} Density profile (heat map) and velocity flow field (arrows, scaled for better visualisation) of the medium from simulations (in the co-moving frame $(u,w)$ scaled by the decay length $\xi$) at $\phi = 0.45$, $T = 0.1$, $f = 2.0$, and $\tau \to \infty$. 
\textbf{B} Corresponding density profile (heat map)  and velocity flow field (arrows, scaled for better visualisation) of the medium obtained from 
the linearised hydrodynamic theory.  \textbf{C} The linearised theory (inset) predicts that the excess density of the passive medium is fore-aft asymmetric and shows an an exponential decay in front  and an algebraic decay behind the moving active particle, which is borne out by the simulations.  The dark solid symbols show averages from multiple simulation runs, together with standard deviation.
\textbf{D-F} The profiles of excess density and velocity components in the moving frame along $u$ at $w=0^{+}$ (red) and $w=0^{-}$ (blue). 
Insets are the results from the linearised hydrodynamic equations. The dark solid symbols show averages from multiple simulation runs, together with standard deviation.}
\label{fig:fig5}
\end{figure*}

\begin{figure*}[t!]
\centering
\includegraphics[width=1\textwidth]{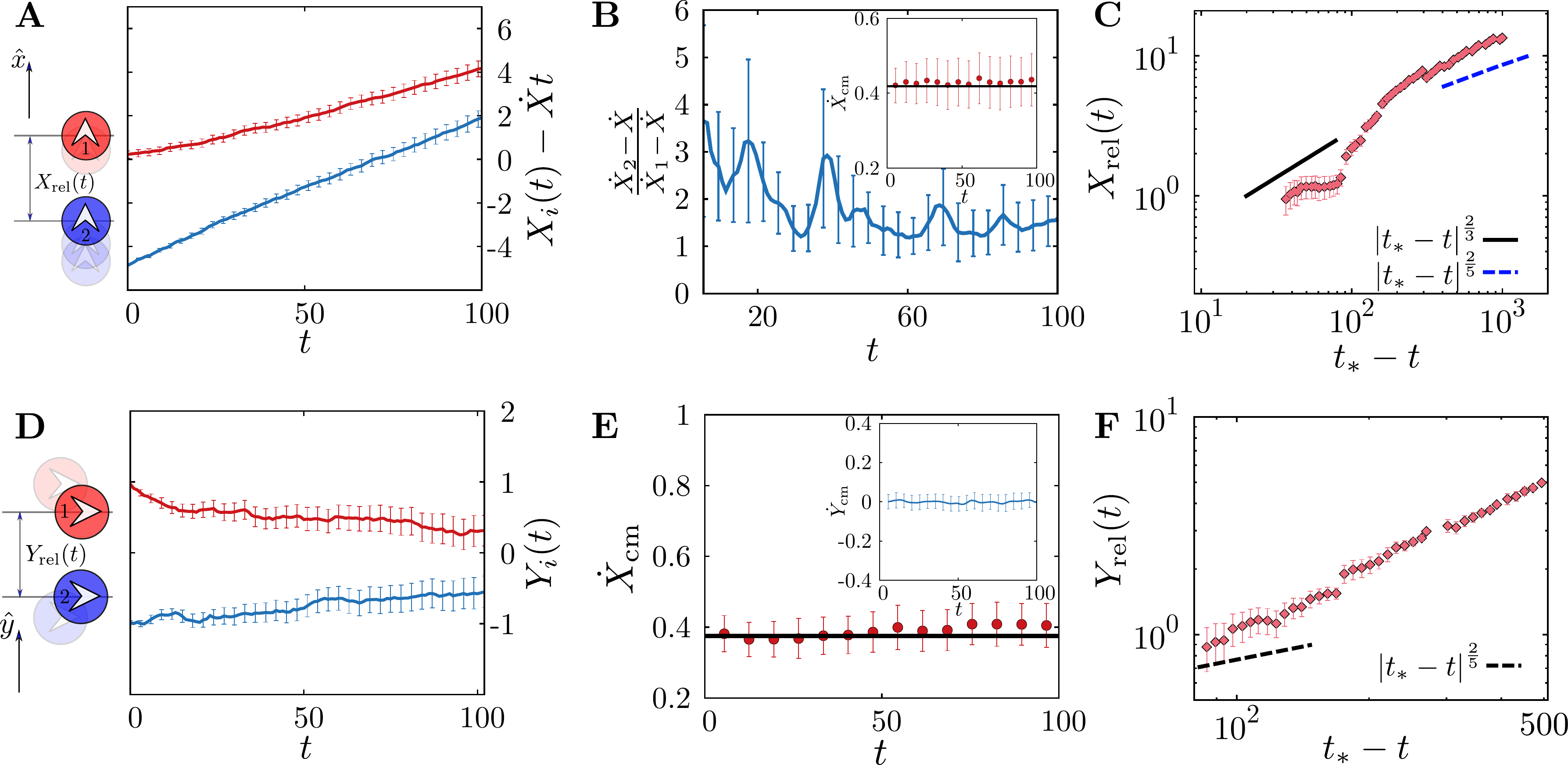}
\caption{\textbf{Dynamics of two particles moving through the passive medium.} 
\textbf{A-C} Two active particles moving along the x-axis (the direction of their active forces whose magnitude $f = 2$, and large persistence time $\tau$) with separation vectors parallel to the direction of motion. The leading particle 1 (red) and trailing particle 2 (blue) are shown to the left. 
\textbf{A} Time dependence of the positions of the pair of active particles (after subtracting the measured single particle displacement $\dot{X} t$), along $\hat{x}$. The particles approach each other as time progresses.
The standard deviation over many initial realisations of the passive particles is shown.
\textbf{B} The ratio of the particle velocities (after subtracting the measured single particle velocity) remains systematically greater than one, which indicates that the trailing particle 2 speeds up to meet the leading particle 1, a manifestation of {\it non-reciprocal interactions}. Inset shows the center of mass velocity $\dot{X}_{\rm cm} (t)$ remains positive and constant throughout (solid line is the prediction from theory). \textbf{C} The two active particles approach each other in finite time $t_*$. The dashed line, a prediction from theory, suggests that the relative position has a scaling form $X_{rel} \propto \vert t - t_{*}\vert^{\alpha}$,
where $\alpha$ crosses over from $2/5$ to $2/3$ as $t \to t_{*}$.   
\textbf{D-F} Two active particles moving along the x-axis with separation vectors perpendicular to the direction of motion (rest same as above).
\textbf{D} Time dependence of $y$-positions of the active particles  showing convergent trajectories. 
\textbf{E} x-component of center of mass velocity is positive and constant, solid line is the prediction from theory. Inset shows y-component center of mass velocity is zero, indicating symmetric approach.
\textbf{F} The two active particles approach each other in finite time $t_*$. The dashed line, a prediction from theory, suggests that the relative position has a scaling form $Y_{rel} \propto \vert t - t_{*}\vert^{\alpha}$,
where $\alpha=2/5$ as $t \to t_{*}$. 
}
\label{fig:fig6}
\end{figure*}

We first look at the dynamics of a single active particle in 
the viscoelastic medium. Let us take the limit of large $\tau$, and so over the time scale of interest, the orientation $\n$ is fixed, say along the $\hat x$ direction. Let us for the moment ignore the back reaction term $C \rho \nabla \rho$ in Eq.\,\ref{eq:activepart1}; we will say that 
$v_0$ is reduced from its bare value $f/\gamma$ in a $\rho$ dependent manner.
Using  Eqs.\,\ref{eq:density}, \ref{eq:forcebalance1} we get,
\be
\pt \rho + B \nabla \cdot \l \rho^2 \nabla \rho\r + f \partial_x \l \rho \delta^{(2)}(x-X(t), y) \r
=0.
\label{eq:anisoburgers}
\ee
This nonlinear equation resembles an anisotropic Burgers equation with a source~\cite{Medina}, and so one might expect travelling pulse solutions. 
We obtain analytical solutions of the linearised equations, by transforming the equation to coordinates in the moving frame of the active particle,
$u = x-X(t)$ and $w=y$, which can then be easily solved using fourier transform methods $(u,w)$ (see {\it Supplementary Information} Sec.~S4). The excess density profile $\rho(u,w)$ in the comoving frame takes the form,
\be
  \label{eq:rhofull}
 \rho(u, w) = \left\{ 
\begin{array}{l}
  2 \mathcal{D}(u) \left[ \mathcal{I}_1(u, w) + \frac{\mathcal{I}_2(u, w)}{\xi} \right] 
 \,\,\,\,\, \mbox{for $u > 0$}
 \\
  2 \mathcal{D}(u) \left[ \mathcal{I}_1(u, w) -  \frac{\mathcal{I}_2(u, w)}{\xi} \right]  \,\,\,\,\,\mbox{for $u < 0$}
\end{array}
\right.
\ee

where,
\bea
\mathcal{I}_1(u, w) & = & \frac{u/\xi}{\sqrt{u^2 + w^2}} K_1 \left[ \frac{\l u^2 + w^2 \r^{\frac{1}{2}}}{\xi} \right] \nn \\
\mathcal{I}_2(u, w) & = & K_0 \left[ \frac{\l u^2 + w^2 \r^{\frac{1}{2}}}{\xi} \right] \nn
\label{eq:I1_I2}
\eea
with $K_0$ and $K_1$ being the modified Bessel functions of the second kind,
and
\be
\mathcal{D}(u) =  \frac{(\rho_0 + \rho(0, 0)) f'}{2 \pi v_0 \xi} \me^{- \frac{u}{\xi}}
\label{eq:Du}
\ee
The decay length $\xi$ is given by,
\be
\xi =  \frac{2 B \rho^2_0} {\Gamma v_0}.
\label{eq:xi}
\ee
Since the fixed direction of motility breaks rotational invariance, it is natural to expect an anisotropy in the density profile. However what comes as a surprise is that moving density profile breaks fore-aft symmetry. 
This is most apparent when we set $w=0^+$,
and use the asymptotic expansion 
$K_{1,0}(z) \sim \sqrt{\f {\pi}{2z}} e^{-z} + \ldots$,
for large $z$. 
We see immediately that in 
the moving frame,
the density profile in front to the motile particle is piled up and decays exponentially over a scale $\xi$ from the pile up. 
The larger the active force $f$, the smaller is $\xi$, implying a sharper pile up.
However, behind the active particle there is a long range underdense region which ``decays'' as a power-law ${\vert u\vert}^{-1/2}$ ({\it Supplementary Information} Sec.~S4).

Knowing the density profile to linear order, we use Eqs.\,\ref{eq:forcebalance1}
and \ref{eq:activepart1} to compute the velocity
flow field of the passive fluid and the velocity of the active particle. A comparison of the density profiles and the velocity flows with the simulation results is shown in Fig.\,\ref{fig:fig5}\textbf{A-F}. The agreement is satisfying; in particular the demonstration in Fig.\,\ref{fig:fig5}\textbf{C} that the excess density profile behind the moving active particle decays as the advertised power-law.
One may in principle, improve on the linear theory by setting up a diagrammatic perturbation expansion. However, since the linear theory compares well
with the numerical simulation of Eq.\,\ref{eq:abp} and with the  ``exact'' numerical solution of the nonlinear equation Eq.\,\ref{eq:anisoburgers} in $d=1$ (next section),
we do not take this up here.

\subsection{Accuracy of linear theory - comparison with ``exact'' numerical analysis of nonlinear equation in $d=1$}

A linear analysis in $d=1$, shows that the density profile is,
\bea
\rho(u) & = & \frac{(\rho_0 + \rho(0, 0)) f'}{ v_0 \xi} \me^{- \frac{u}{\xi}}\,\,\,\, \mbox{for $u>0$} \nn\\
& = & 0\,\,\,\, \mbox{for $u<0$}\,, 
\label{eq:1dprofile}
\eea
where the length scale is given by $\xi=B \rho_0^2/v_0$ ({\textit{Supplementary Information}} Sec.~S5).
The density piles up in front of the active particle and decays exponentially ahead of it, while behind the active particle there is no wake. We now check to see how this calculated profile compares with an exact numerical solution of the nonlinear equation Eq.\,\ref{eq:anisoburgers}. The accurate numerical solution of this nonlinear PDE requires some care due to shock forming tendencies in the convective term ({\textit{Supplementary Information}} Sec.~S6). The result for the density profile of the travelling pulse is shown in {\it Supplementary Fig.~S5}. The comparison with the linear theory is quite good, the absence of the wake is vividly apparent in the 1d exact numerical solution ({\it Supplementary Fig.~S5}).

\subsection{Two active particles moving through the compressible medium}
The fore-aft asymmetric long range density wake around the motile particle has an unusual effect on the interactions between two or more motile particles. This is best illustrated by considering the dynamics of two motile particles in a simplifying geometry where both particles move in the same direction with their separation vector being parallel or perpendicular to the direction of motion.

Let the trajectories of the two active particles be represented by
 ${\R}_1(t), {\R}_2(t)$, in the limit of large persistence time, so that we can take 
the orientations $\n_1$ and $\n_2$ to be time independent.
 To $\mathcal{O}(\delta \rho)$, these particles, individually, leave a time dependent anisotropic and fore-aft asymmetric wake described by $\rho(x,y, t)$, whose
back-reaction on the movement of the active particles themselves, is easily estimated 
\bea
\label{eq:twopart1}
 \gamma {\dot \R}_1 & = & f \n_1 -  C \rho_0 \nabla \rho \Big \vert_{\mbox{self}_1}  - C \rho_0 \nabla \rho \Big \vert_{1 \leftarrow 2} \\
\gamma {\dot \R}_2 & = & f \n_2 - C \rho_0 \nabla \rho \Big \vert_{\mbox{self}_2} - C \rho_0 \nabla \rho \Big 
\vert_{2 \leftarrow 1} 
\label{eq:twopart2}
\eea
where ${i \leftarrow j}$ denotes the effect of particle $j$ on particle $i$. 
These equations may be cast in terms of the relative coordinate $\R_{rel} = \R_1-\R_2$ and the centre of mass $\R_{cm} = (\R_1+\R_2)/2$. For the two geometries under consideration, we find the following (details in {\textit{Supplementary Information}} Sec.~S4) --\\

\begin{enumerate}
    \item The separation vector between the leading particle 1 and the trailing particle 2 is along $\hat{x}$ direction, parallel to their direction of motion (Fig.\,\ref{fig:fig6}\textbf{A}).

    We find that while the centre of mass velocity ${\dot X}_{cm}>0$ (Fig.\,\ref{fig:fig6}\textbf{B} inset), the interparticle separation $X_{rel}$ {\it decreases} in time (Fig.\,\ref{fig:fig6}\textbf{A}),
    i.e. ${\dot X}_{rel}<0$, starting from the initial value.  More interestingly, the speed of approach of the particle 2 to particle 1, {\it increases} as $X_{rel} \to 0$.
    This nonreciprocal sensing~\cite{SRAS2020,SSSR2014,KHMR}, is a consequence of the fore-aft asymmetric wake and causes the trailing particle to catch up with the leading one in a {\it finite time} 
    (Fig.\,\ref{fig:fig6}\textbf{B}). 
    
    A dominant balance analysis of the equation for $X_{rel}$ shows a crossover scaling of the form $X_{rel} \propto \vert t-t_{*}\vert^{\alpha}$ where $\alpha$ goes from $2/5$ to $2/3$ as the particles approach each other (Fig.\,\ref{fig:fig6}\textbf{C}).

     \item The separation vector between the particle 1 and the particle 2 is along $\hat{y}$ direction, perpendicular to their direction of motion (Fig.\,\ref{fig:fig6}\textbf{D}). 
   
    Here again the centre of mass velocity ${\dot X}_{cm}>0$ (and is the same as the single particle speed,
    Fig.\,\ref{fig:fig6}\textbf{E}), and the inter-particle separation $Y_{rel}$ {\it decreases} in time (Fig.\,\ref{fig:fig6}\textbf{D}),
    i.e. ${\dot Y}_{rel}<0$, starting from the initial value.  This leads to the  trajectories of particles 1 and 2  converging towards each other in a symmetrical manner (Fig.\,\ref{fig:fig6}\textbf{D}). 
    
    An asymptotic analysis shows that the relative position $Y_{rel} \propto \vert t-t_{*}\vert^{2/5}$ as $t \to t_{*}$ (Fig.\,\ref{fig:fig6}\textbf{F}).
    The approach of the two particles is slower than in the previous case.

\end{enumerate}

The second case resembles the magnetic force between two parallel wires carrying current in the same direction, and is a consequence of the breaking of time reversal symmetry. Likewise, a pair of active particles initially moving towards each other, will scatter off (repel) and will slow down as they move away from each other.
The scattering of active particles in other geometries can also be worked out to this order. 

\section{Discussion}

In this paper, we have studied the dynamics of a dilute suspension of active Brownian 
particles moving through a dense compressible passive fluid that dissipates momentum through friction. 
The dynamical  interplay between the active particles and the passive medium, not only results in a remodelling of the passive medium, but also in a 
back-reaction on the movement of the active particles themselves. 
Such active {\it ploughers} show a jamming transition at fixed density of the medium. 
In the unjammed phase, a moving active plougher generates a fore-aft asymmetric density wake
which is the source of the long-range nonreciprocal interaction between {\it moving} active particles mediated dynamically through the passive compressible medium.  This emergent nonreciprocal interaction is a consequence of a dynamical phase transition to a state with finite current. This leads to a nonreciprocal sensing wherein a trailing particle senses and catches up with a leading particle moving ahead of it. Further, the movement of the active particle leaves a dynamical trace on the responsive medium, these effects of nonreciprocity are more 
indelibly manifest in the vicinity of the jamming-unjamming transition.


We recall that in Eq.\,\ref{eq:activepart2}  we have assumed that the direction of the propulsion force is set by some mechanism internal to the active particle and   therefore
independent of the passive medium. 
To experience the full scope of nonreciprocal effects possible here, one needs to extend the hydrodynamic equations Eqs.\,\ref{eq:activepart1}, \ref{eq:activepart2},
to include an active torque that drives
 $\n_i$ to align along the direction of the smallest (largest) density gradient $\nabla \rho$ -- this will lead to both {\it taxis} and {\it phoresis}, features that have been explored in~\cite{SSSR2014, KHMR} in other contexts. Such considerations lead to a simple physical version of sense-and-capture, even in the absence of any kind of chemical sensing.

The success of our hydrodynamic analysis motivates us to go beyond the study of one and two active  particles and look at many-body effects.  In~\cite{Mandal2016}, we had seen how the minority component self-propelled particles cluster on account of activity;  the long range
nonreciprocal interaction observed here, will translate to a new kind of {\it non-reciprocal motility induced clustering}~\cite{Cates2015} of active particles mediated by the passive medium. This and its relationship with the anisotropic Burgers equation with coloured noise~\cite{Medina} will be taken up later.

\section{acknowledgments}
JPB and RM contributed equally to this work. We acknowledge support from the Department of Atomic Energy (India), under project no.\,RTI4006, and the Simons Foundation (Grant No.\,287975), and computational facilities at NCBS. RM acknowledges funding from the European Union’s Horizon 2020 research and innovation programme under Marie Sklodowska-Curie grant agreement No.\,893128.
ST acknowledges funding from the Human Frontier Science Program and the Max Planck Society through a Max-Planck-Partner-Group.
MR acknowledges a JC Bose Fellowship from DST-SERB (India).

\end{document}


\begin{center}
{\Large Supplementary Information}
\vskip 0.6in
{\Large \bf{Active ploughing through a compressible viscoelastic fluid}}
\vskip 0.2in

\author{Jyoti Prasad Banerjee}
\affiliation{Simons Centre for the Study of Living Machines, National Centre for Biological Sciences (TIFR), Bangalore, India}
\author{Rituparno Mandal}
\affiliation{Institute for Theoretical Physics, Georg-August-Universit\"at G\"ottingen, 37077 G\"ottingen, Germany }
\author{Deb Sankar Banerjee}
\affiliation{Department of Physics, Carnegie Mellon University, Pittsburgh, USA}
\author{Shashi Thutupalli} 
\affiliation{Simons Centre for the Study of Living Machines, National Centre for Biological Sciences (TIFR), Bangalore, India}
\affiliation{International Centre for Theoretical Sciences (TIFR), Bangalore, India}
\author{Madan Rao}
\affiliation{Simons Centre for the Study of Living Machines, National Centre for Biological Sciences (TIFR), Bangalore, India}

\maketitle

\end{center}
\tableofcontents

\clearpage
\newpage

\section{Agent-based simulations}
\label{sec:simmodel}

To ensure that the assembly of soft particles of area fraction $\phi$ remains in a positionally disordered state, we work with a modified 2D Kob-Andersen $65:35$ (A:B) binary mixture~\cite{Kob,Bruning2008}, a prototype glass-forming liquid. Particles $i$ and $j$, interact
via a 2-body soft repulsive potential,
\begin{equation}\label{eq:pot}
    V_{ij} =  
\begin{cases}
    4\epsilon_{ij} \l\frac{\sigma_{ij}}{r_{ij}}\r^{12} + v_{0} + v_{2}\l\frac{\sigma_{ij}}{r_{ij}}\r^{-2}
    + v_{4} \l\frac{\sigma_{ij}}{r_{ij}}\r^{-4},
        & \text{if } r_{ij} < r_{c, ij}\\
    0,              & \text{otherwise}
\end{cases}
\end{equation}
where $r_{ij}$ is the distance between the particles $i$ and $j$. Our choice of parameter values,
$v_0 = -112 \epsilon_{ij}$, $v_2 = 192 \epsilon_{ij}$, $v_4 = -84 \epsilon_{ij}$ and the cut-off
$r_{c, ij}= \sigma_{ij}$,  ensures that both the potential and the force is smooth at the cut-off distance and the interaction is purely repulsive.
We fix the energy and length scales to be in the units of $\epsilon_{AA}$ and $\sigma_{AA}$, respectively by setting $\epsilon_{AA} = 1.00$, $\sigma_{AA} = 1.00$. The other parameters are chosen following the Kob-Andersen model~\cite{Kob} {\it{i.e.}} $\epsilon_{BB} = 0.50$, $\sigma_{BB} = 0.88$, $\epsilon_{AB} = 1.50$, $\sigma_{AB} = 0.80$.

Of these a small fraction of particles $\phi_a$ is made active - their dynamics is described by active Brownian particles (ABP)~\cite{RMP2013, Fily2012, Takatori2015, Codina2017, Cates2015, Rituparno2016} immersed in a background of passive particles. All particles 
are subject to a thermal noise $\vartheta$ of zero mean and variance equal to $2 \gamma T$ (setting
$k_B=1$), obeying FDT. The subset $i \in \mathcal{A}$ of ABPs are subject to
additional active stochastic forces ${\bf f}_i = f\, \mathbf{n}_i \equiv f\,\l \cos \theta_i, \sin \theta_i\r$.
The orientation of the propulsion force ${\theta}_i$ undergoes rotational diffusion, described by an
athermal noise $\xi_i$, with  zero mean and correlation $\langle \xi_i(t) \xi_j(t') \rangle = 2 \tau^{-1} \delta_{ij} \delta(t-t')$. 
Its effect on the particle-dynamics appears as an exponentially correlated vectorial noise with correlation time $\tau$,
which being unrelated to the drag $\gamma$, is not constrained by fluctuation-dissipation relation.

The full dynamics is described by the  
Langevin equation,
\bea
m{\ddot{\mathbf{x}}}_i & = & -\gamma {\dot{\mathbf{x}}}_i - \sum_{i\neq j}^{N} \pa_j V_{ij} + f \mathbf{n}_i
\,\mathbbm{1}_{(i\in\mathcal{A})}+ \vartheta_i\,,  \nn\\
\dot{\theta}_i & = & \xi_i\,\,\,\,\,\,\,\,\, \mbox{for} \,\,{i\in\mathcal{A}}\,.
\label{eq:abp}
\eea
where $\mathbbm{1}_{(i\in\mathcal{A})}$ is the indicator function which ensures that the active forces are only imposed on particles $i$ belonging to the active set $\mathcal{A}$.

To proceed with the numerical simulation, we take the overdamped limit (drop inertia) and convert the resulting dynamical equation to non-dimensionless form by scaling,
\be
t \to \f{\gamma}{\epsilon_{AA}} t, \,\,\, {\bf x}_i \to  \sigma_{AA} {\bf x}_i, \,\,\,
 V_{ij} \to  \epsilon_{AA} V_{ij}, \,\,\,
T \to  \epsilon_{AA} T
\ee
and setting $\sigma_{AA} = \epsilon_{AA} = 1$, to arrive at,

\bea
\frac{dx_i}{dt} & = & - \sum_{i\neq j}^{N} \frac{\pa}{\pa x_j} V_{ij} + f (\mathbf{n}_i \mathbf{.} \mathbf{\hat{x}}_i) \,\mathbbm{1}_{(i\in\mathcal{A})} + \sqrt{\frac{2 T}{dt}} \zeta_i\
\label{eq:abp4}
\eea
where $\langle \zeta_i(t)\rangle = 0$ and $\langle \zeta_i(t) \zeta_j(t') \rangle = \delta_{ij} \delta(t-t')$.
Similarly, the orientational dynamics (which is purely diffusive), can be written as,
\bea
\dot{\theta_i} & = & \sqrt{\frac{2}{\tau}} \eta_i
\label{eq:abp5}
\eea
where, $\langle \eta_i(t)\rangle = 0$ and $\langle \eta_i(t) \eta_j(t') \rangle = \delta_{ij} \delta(t-t')$.

We perform Brownian dynamics (BD) simulations at fixed particle-number, volume of the system and temperature of the heat-bath (NVT) in 2-dimensions using a square box of reduced length $L = 45$, with periodic boundary conditions (PBC). We use predictor-corrector algorithm (Euler-Trapezoidal method) and forward Euler method to update the positions and orientations, respectively~\cite{Butcher, Frenkel}. For all the simulations related to Figs.\,1, 2 (main text) we keep the area fraction of active particles fixed at $\phi_a = 0.017$ (dilute limit), and vary the number of passive particles constituting the medium to control the overall density or area fraction $\phi$. The simulations to generate data for Figs.\,3, 4, 5 (main text) are performed with one active particle in the passive medium.

\section{Characterisation of approach to glass transition}
\label{sec:glass}

\begin{figure*}[h]
\centering
\includegraphics[width=0.95\textwidth]{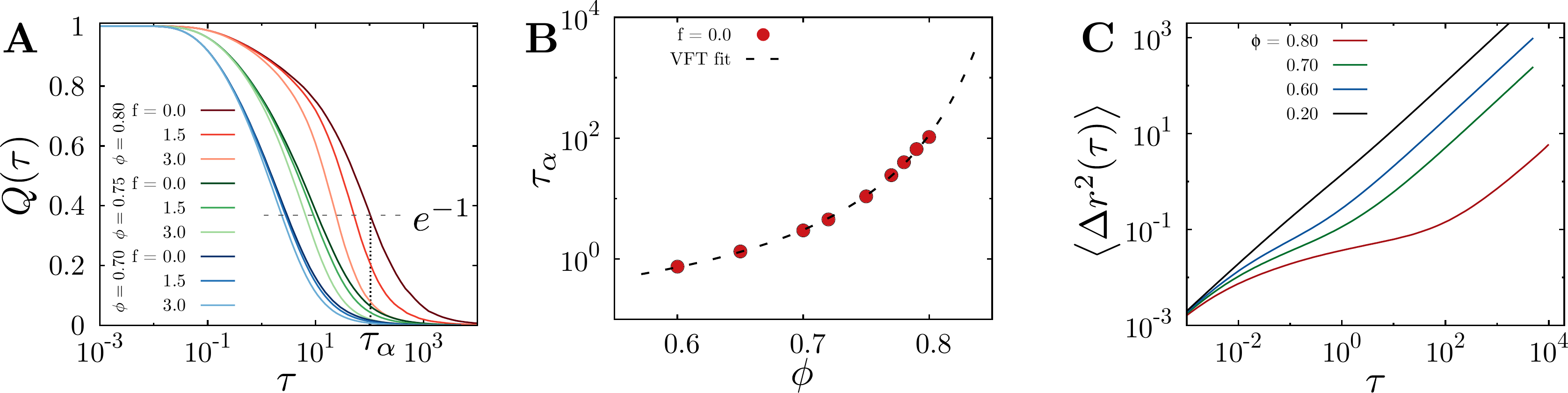}
\caption{{\bf A} Q($\tau$) for three different area fractions of the medium and three different values of active force at $T = 0.5$. The gray dashed line represents $Q(\tau = \tau_{\alpha}, \phi, T) = e^{-1}$ as in Eq.~\ref{eq:tau_alpha}. The corresponding $\tau$ is $\tau_{\alpha}$, as shown by the black dotted line. 
{\bf B} VFT fit of $\tau_{\alpha}$ vs. area fraction ($\phi$) for the passive medium ($f = 0.0$).
{\bf C} Mean squared displacements of the medium at different area fractions. The lowest area fraction ($\phi = 0.20$) represents the weak interaction limit, while the highest one ($\phi = 0.80$) shows stong caging effect upto $\tau \approx 10^{2}$.
}
\label{fig:SI01}
\end{figure*}

For different values of the area fraction $\phi$, we first equilibrate the passive medium at the reduced temperature $T = 0.5$. 
We then monitor the approach of the passive medium towards a glass transition, by computing the relaxation in its density fluctuations, measured by the two point overlap correlation function (Fig.\,\ref{fig:SI01}{\bf A}); the transport of tagged particles, measured by the mean squared displacement (Fig.\,\ref{fig:SI01}{\bf C}) and its rheological properties, as we systematically vary $\phi$. Only in this section, $\tau$ indicates the time elapsed starting from the time-origin, $t$, and is not to be confused with the $\tau$ used in all the other sections of this {\it Supplementary Information} as well as in the main text, representing the persistence time of the active particles. 
\\


\noindent
{\bf Overlap function}, $Q(\tau)$, a measure of the density relaxation, is defined as 
\begin{equation} \label{eq:Qt}
 Q(\tau; \phi, T) = \Big \langle \frac{1}{N} \sum_{i = 1}^{N} \omega \l \vert \mathbf{x}_{i}(t+\tau) - \mathbf{x}_{i}(t)\vert \r \Big \rangle
\end{equation}
with $\omega(x) = 1$ for $x< 0.3$ and zero otherwise. 
The $\left <...\right>$ sign indicates averaging over different time-origins ($t$), and different realisations of the system. 
The $\alpha$-relaxation time, $\tau_{\alpha}$ is calculated 
by setting 
\begin{equation} \label{eq:tau_alpha}
  Q(\tau = \tau_{\alpha}; \phi, T) = \frac{1}{e}
 \end{equation}
Then $\tau_{\alpha}$, for different area fractions $\phi$, is  fit to the Vogel-Fulcher–Tammann (VFT) form (Fig.\,\ref{fig:SI01}{\bf B}),
 \be{}
 \label{eq:VFT}
 \tau_{\alpha} = \tau_{\infty} \exp \l \frac{B}{\phi_{\rm VFT} - \phi} \r
 \ee{}
to obtain $\phi_{\rm VFT}$, the VFT glass transition area fraction.
\\

\noindent
{\bf Mean squared displacement (MSD)}, is defined as,
\be 
\label{eq:MSD}
\Big  \langle \Delta r^2 (\tau) \Big \rangle = \Big \langle \frac{1}{N} \sum_{i = 1}^{N} \left |\mathbf{x}_{i}(t+\tau) - \mathbf{x}_{i}(t) \right |^2 \Big \rangle
\ee{}

The $\langle \ldots \rangle$ denotes averaging over the time-origins and multiple statistically independent realizations. The long time diffusion coefficient ($D_{\infty}$) is computed from the long time limit of the MSD,
\be{} \label{eq:Dinfty}
D_{\infty} = \frac{1}{4 \tau} \lim_{\tau \to \infty} \Big \langle \Delta r^2 (\tau) \Big \rangle
\ee{}


\noindent
{\bf Passive microrheology}, is a measure of the viscoelastic(shear) properties of the medium, and can be  obtained from the frequency transformed MSD,
$\langle \Delta r^2 ({\omega}) \rangle$~\cite{Mason},
\bea
\label{eq:moduli}
G^{\prime}(\omega) & = & \vert G^*(\omega)\vert  \cos{(\frac{\pi \alpha(\omega)}{2})} \nn \\
G^{\prime\prime}(\omega) & = & \vert G^*(\omega)\vert \sin{(\frac{\pi \alpha(\omega)}{2})}
\eea

where, 

\be
\label{eq:complexModuli}
\vert G^*(\omega)\vert \approx \frac{K_B T}{\pi a \langle \Delta r^2 (\omega) \rangle \Gamma\left(1 + \alpha(\omega)\right)}
\ee

\be{}
\label{eq:alphaOmega}
\alpha(\omega) = \frac{d \ln{\langle \Delta r^2 (\tau) \rangle}}{d \ln{\tau}} \bigg|_{\tau = \frac{1}{\omega}}
\ee{}
Here, `a' is the radius of the particle and $\Gamma(x)$ is the $\Gamma$-function.

We choose $T = 0.5$ for all the simulations related to Fig.\,1 as 
(1) at this temperature cage-hopping dynamics is clearly visible for higher area fractions (i.e., $\phi = 0.7-0.8$) of the medium, and
(2) for moderate area fractions (i.e., $\phi = 0.4-0.6$) the temperature is not too close to the glass transition temperature where the dynamics becomes extremely slow and it becomes very difficult to acquire enough data within a reasonable run-time. 
Both the equilibration and data-generation runs are for a time $\geq$ 100 $\tau_{\alpha}$.


\section{Morphology of underdense regions in the medium excavated by the active particles}
\label{sec:morph}

\subsection{Creation of a density map from agent-based simulations}
\label{subsec:densmap}

We use the instantaneous configuration (i.e., position-coordinates of all the passive particles) to generate a coarse-grained density field that has been used in Figs.\,3-5 in the main text. For this, we divide the entire simulation box of size $L_0$ into a square grid of grid length $L_{g}$, and number of cells $N_g = (\frac{L_0}{L_{g}})^2$. For a typical cell $(k, l)$-th cell, we compute the density field using the following formulae
\be{}
\label{eq:rhocg}
\rho (k, l) = \frac{1}{\mathcal{N}} \sum_{i = 1}^N \exp{\left ( -\frac{d}{d_0} \right )}
\ee{}
where $d = | \mathbf{x}_i - \mathbf{x}_{k,l} |$ is the distance of the i-th particle from the centre of the grid ($\mathbf{x}_{k,l}$) and $d_0$ is the coarse-graining length scale, and $\mathcal{N}$ is the normalisation constant. We have used different values for $L_g$ and $d_0$, ensuring that $d_0 > L_g$ and that both are $\sim \mathcal{O}(1)$. For Fig.\,3, since we compare the density fields between different area fractions, we choose
\be{}
\mathcal{N} = \sum_{k, l = 1}^{\sqrt{N_g}} \sum_{i = 1}^N \exp{\left ( -\frac{d}{d_0} \right )}
\ee{}
which ensures that Eq.~\ref{eq:rhocg} gives us the normalised probability density of the passive medium. For Fig.\,4 and Fig.\,5 (main text), we choose
\be{}
\mathcal{N} = \frac{1}{N} \sum_{k, l = 1}^{\sqrt{N_g}} \sum_{i = 1}^N \exp{\left ( -\frac{d}{d_0} \right )}
\ee{}
which ensures that Eq.~\ref{eq:rhocg} gives us normalised number density.

\subsection{Identification of under-dense regions and analysis of their geometry}

We use the density profile that we discussed 
in Sect.\,\ref{subsec:densmap} and identify the low density regions by employing an upper cut-off ($\sim 4 \times 10^{-4}$) on the grid defined above. The results described in Fig.\,3{\bf B} in main text are robust to reasonable variation of this threshold.  Any grid point (cell) that has density less than or equal to the aforementioned threshold is considered as low density region. This low density regions are represented by a set of points defined on the grid ${\mathbf z}_i$, where $i \in [1, N_u]$ where $N_u$ is the total number of under-dense grid points. We then compute moment of gyration tensor, $M_{\alpha \beta} = \sum_{i = 1}^{N_u} (z_{i}^{\alpha} - z_{\rm{cm}}^{\alpha})(z_{i}^{\beta} - z_{\rm{cm}}^{\beta})$, to characterise the spatial distribution or structures of such point where  ${\mathbf z}_{\rm cm} = \frac{1}{N_u} \sum_{i = 1}^{N_u} {\mathbf z}_i$.

\begin{figure*}[h]
\centering
\includegraphics[width=0.95\textwidth]{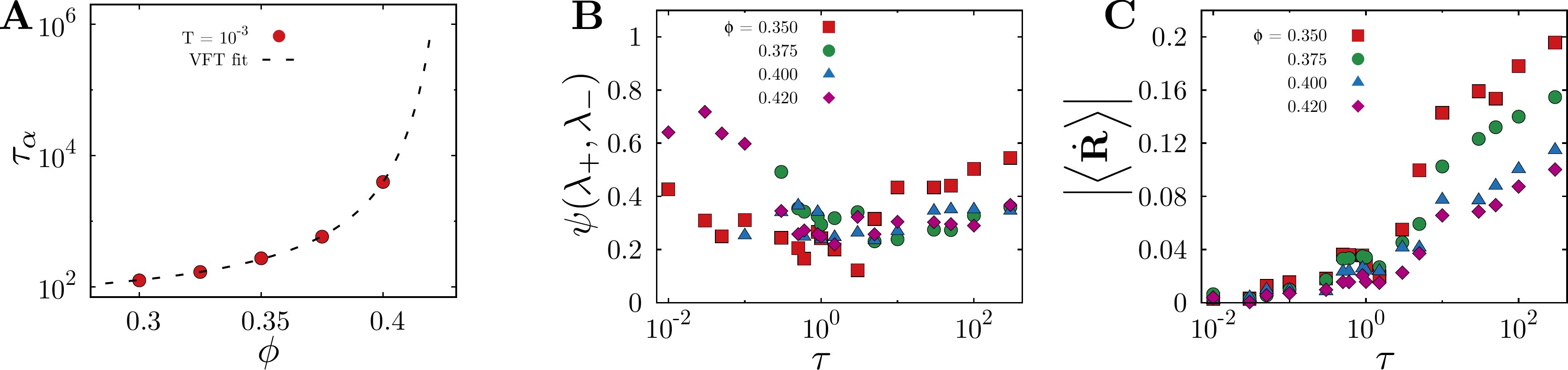}
\caption{{\bf A} $\alpha$-timescale of the passive medium as a function of the area fraction $\phi$ and the corresponding VFT fit. The value of $\tau_{\alpha}$ for $\phi$ = 0.420 in Fig.\,3{\bf A} is taken from the VFT fit.  
{\bf B} Shape parameter (defined in Fig.\,2 in main text) as a function of the persistence time $\tau$ of the active particle for small active forcing ($f = 0.5$). 
{\bf C} Magnitude of the averaged velocity of the active particle with $f = 0.5$ in a passive medium of varying area fraction, as a function of $\tau$.}
\label{fig:SI03}
\end{figure*}


%
%
%



\section{Linearised hydrodynamic theory}
\label{sec:lineartheory}

We first write the full dynamics of the medium and the active particles, and their interplay.
The medium is described by a local density field 
$\rho(\br,t)$ and a velocity field $\v(\br,t)$, and obeys,
the continuity equation for $\rho$
\be
\pt \rho + \nabla \cdot \l \rho \v \r = 0
\label{eq:density}
\ee
and force-balance equation for $\v$,
\be
{\mathbf \Gamma} \v = - B \rho \nabla \rho + f \,\sum_i \n_i(t)\, \delta^{(2)}({\bf r}- \R_i(t)) \label{eq:forcebalance}
\ee
In Eq.\,\ref{eq:forcebalance} above, $B>0$, since the pressure due to the medium opposes the local forcing from the active particles.
The dynamics of the active particles in the dilute limit is given by 
\bea
\gamma \dot \R_i & = & \,f \n_i - C \rho \nabla \rho \Big \vert_{at\,\, i} \\
\dot \n_i & = & \mbox{{\boldmath $\xi$}}_i(t)
\label{eq:activepart}
\eea
where $\gamma$ is the drag on the active particle and $C$ is the extent of back reaction of medium on the translational dynamics of the active particle. The vector orientational noise {\boldmath $\xi$}$_i$ has zero mean and is exponentially correlated, with correlation time $\tau$.


In general, the frictions ${\mathbf \Gamma}$, $\gamma$, the active force magnitude  $f$ and coefficients $B$ and $C$ could depend on $\rho$.

\subsection{Single active particle moving through the compressible medium in $d=2$}
\label{sec:linearone}

We will now restrict ourselves to a single active particle moving through the compressible medium,
\be
{\mathbf \Gamma} \v = - B \rho \nabla \rho + f \,\n(t) \,\delta^{(2)}({\bf r}- \R(t)) 
\label{eq:forcebalance1}
\ee
where $\R(t)$ is the coordinate of the active particle.
As the active particle moves through the medium, it creates density inhomogeneities, which relax over time; we will be interested in a linear analysis about the uniform density of the ploughed medium. In this limit, we take ${\mathbf \Gamma}$ and $B$ to be independent of $\rho$, and ignore the back-reaction term $C$. 

 Plugging Eq.\,\ref{eq:forcebalance1} into the continuity equation Eq.\,\ref{eq:density}, we get
\be
\pt \rho - B \nabla \cdot \l \rho^2 \nabla \rho  \r 
 + f \,\nabla \cdot \l \rho \n \,\delta^{(2)}({\bf r}- \R(t))  \r = 0
\label{eq:rhoevo2}
\ee
where we have absorbed $\Gamma$ into a redefinition of $B$ and $f$.
To simplify analysis, we will assume that the persistence time of the active particle is large compared to the timescale of observation, i.e., 
 $\n(t)$ is independent of time.  After initial transients, the density of the 
medium can then be written in terms of the collective coordinate,
\be
\rho({\bf r}, t) = \rho({\bf r} - {\bf R}(t)) \nn 
\ee
Transforming coordinates 
to the co-moving frame of the active particle (without loss of generality, we take  $\n$ to point along the ${\hat{\x}}$-direction),
\bea
u & = & x - X(t) \nn \\
w & = & y
\label{eq:transform}
\eea
and so,
\bea
\pt & \longrightarrow & \f{\partial u}{\partial t} \pu = - \dot X(t) \pu = -v_0 \pu \nn \\
\partial_x & \longrightarrow & \f{\partial u}{\partial x} \pu = \pu \nn \\
\partial_y & \longrightarrow & \f{\partial w}{\partial y} \pw = \pw
\label{eq:transformedDeriv}
\eea
where $v_0 \equiv f/\gamma$. The equation for the transformed density $\rho(u,w)$ in the co-moving frame now reads,

\be
- v_0\, \pu \rho - B\, \nabla \cdot \l \rho^2 \nabla \rho \r + 
f \, \pu \l  \rho \delta^{(2)}(u, w)\r = 0
\label{eq:transformedEq}
\ee
where $\nabla \equiv (\pu, \pw)$.
\\

We now express the local density as a small deviation (positive or negative) from the uniform $\rho_0$, and rewrite the above equation to linear order in this small deviation,
i.e., we set $\rho \to \rho_0 + \rho$ in Eq.\,\ref{eq:transformedEq}, and linearise to get,

\be
- v_0 \, \pu \rho - B \,\rho^2_0 \, \nabla^2 \rho + f\, \pu \l (\rho_0 + \rho) \,\delta^{(2)}(u, w) \r = 0
\label{eq:linearisedEq}
\ee

This linear equation can be solved by fourier transforming $(u,w) \rightarrow (q_u, q_w)$, and then evaluating the inverse fourier transform using contour integration.
To proceed, we first note that, 
\be{}
\intpminf \intpminf (\rho_0 + \rho(u, w))\, \delta^{(2)}(u, w)\, \me^{-i q_u u} \me^{-i q_w w} \,du \,dw  = \rho_0 + \rho(0, 0)
\label{eq:defFourier}
\ee{}
which leads to,
\be{}
-i v_0 q_u \fou{\rho} + B \rho^2_0 q^2 \fou{\rho} + i f q_u (\rho_0 + \rho(0, 0)) = 0
\label{eq:eqFourier}
\ee{}
where $q^2 = q^2_u + q^2_w$, from which we get,
\be{}
\fou{\rho}(q_u, q_w) = \frac{i f q_u (\rho_0 + \rho(0, 0))}{i v_0 q_u - B \rho^2_0 q^2}\,.
\label{eq:solnFourier}
\ee{}

We now obtain $\rho(u, w)$, by inverse fourier transforming $\fou{\rho}(q_u, q_w)$,
\bea{}
\rho(u, w) & = & \frac{1}{4 \pi^2} \intpminf \intpminf \fou{\rho}(q_u, q_w) \,\me^{i q_u u} \, \me^{i q_w w} \,dq_u \,dq_w \nn \\
& = & - \frac{i f (\rho_0 + \rho(0, 0))}{4 \pi^2 B \rho^2_0} \intpminf \intpminf
\frac{q_u}{q^2_u - \frac{i v_0}{B \rho^2_0}q_u + q^2_w} \me^{i q_u u} \me^{i q_w w} \,dq_u \,dq_w \\
& = & \intpminf I(u, q_w) \me^{i q_w w} \,dq_w
\label{eq:defInvFourier}
\eea{}
where $I$ is an integral over $q_u$, keeping $q_w$ fixed,
\be{}
I(u, q_w) = - \frac{i f (\rho_0 + \rho(0, 0))}{4 \pi^2 B' \rho^2_0} \intpminf \frac{q_u}{q^2_u - \frac{i v_0}{B \rho^2_0}q_u + q^2_w} \me^{i q_u u} \,dq_u
\label{eq:rho_u_qw}
\ee{}
 

We evaluate $I$ in \eqref{eq:rho_u_qw} using contour integration. For this, we determine the simple poles $q_{\pm}$ of the integrand by factorization,
\be{}
q_{\pm}  
 = i \left( \xi^{-1} \pm \sqrt{ \xi^{-2}  + q^2_w}\right)
\label{eq:qpm}
\ee{}
where the length-scale $\xi$ is given by,
\be{}
\xi =  \frac{2 B \rho^2_0}{v_0} \equiv \frac{2 B \gamma \rho^2_0}{f \Gamma}
\label{eq:xi}
\ee{}
when we restore the original parameter definitions.


Note that the sgn\,$\operatorname{Im}(q_+) = +1$ and sgn\,$\operatorname{Im}(q_-) = -1$.
Thus we close the contour of  integration in the upper half plane around $q_+$ when $u > 0$, and in the lower half plane around $q_-$ when $u < 0$.

\begin{figure*}[h]
\centering
\includegraphics[width=0.95\textwidth]{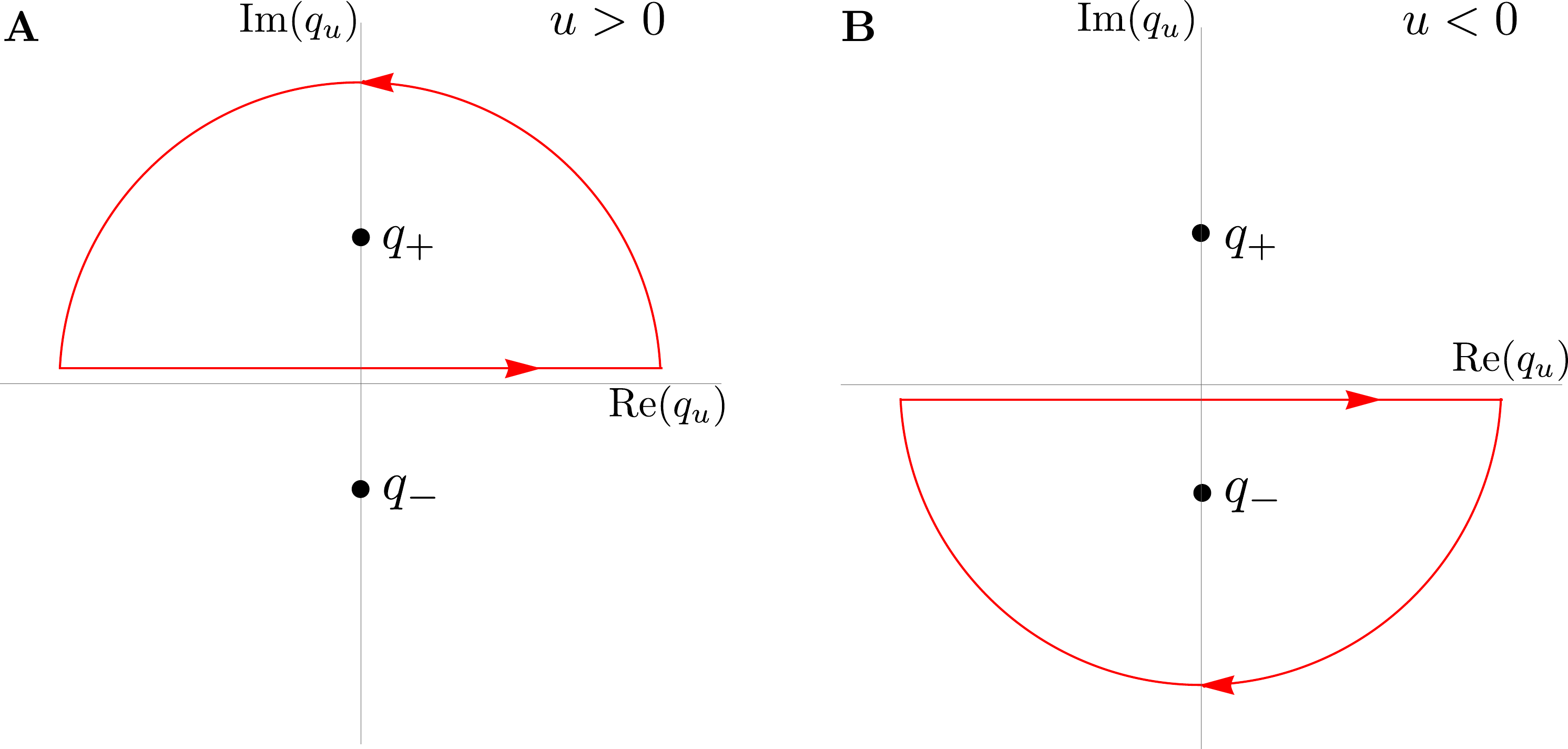}
\caption{ Choice of contours for {\bf A} $u>0$ and {\bf B} $u<0$.  
}
  \label{fig:figSI1}
\end{figure*}

Using, 
$q_+ - q_- 
= 2 i  \sqrt{ \xi^{-2}  + q^2_w}$, 
we can evaluate the residues,
\bea{}
Res(q_+) & = & \frac{q_+}{q_+ - q_-} \me^{i q_+ u} \nn \\
& = & \frac{1}{2} \left[1+ \frac{\xi^{-1}}{\sqrt{\xi^{-2}  + q^2_w}}\right]
 \me^{- \left( \xi^{-1} + \sqrt{ \xi^{-2}  + q^2_w}\right) u}
\label{eq:resq+}
\eea{}
and 
\bea{}
Res(q_-) & = & \frac{q_-}{q_- - q_+} \me^{i q_- u} \nn \\
& = & \frac{1}{2} \left[1- \frac{\xi^{-1}}{\sqrt{\xi^{-2}  + q^2_w}}\right]
 \me^{- \left( \xi^{-1} - \sqrt{ \xi^{-2}  + q^2_w}\right) u}
\label{eq:resq-}
\eea{}
The integral $I$ \eqref{eq:rho_u_qw} can now be read out.
For $u > 0$,
\bea{}
I(u, q_w) & = & - \frac{i f (\rho_0 + \rho(0, 0))}{4 \pi^2 B \rho^2_0} \,2 \pi i \,Res(q_+) \nn \\
& = & \frac{f (\rho_0 + \rho(0, 0))}{4 \pi B \rho^2_0} 
\left[1+ \frac{\xi^{-1}}{\sqrt{\xi^{-2}  + q^2_w}}\right]
 \me^{- \frac{u}{\xi}}  \me^{- u \sqrt{ \xi^{-2}  + q^2_w} }\\
\label{eq:I+}
\eea{}
For $u < 0$,
\bea{}
I(u, q_w) & = & - \frac{i f (\rho_0 + \rho(0, 0))}{4 \pi^2 B \rho^2_0} \,2 \pi i \,Res(q_-) \nn \\
 & = & \frac{f (\rho_0 + \rho(0, 0))}{4 \pi B \rho^2_0} 
\left[1- \frac{\xi^{-1}}{\sqrt{\xi^{-2}  + q^2_w}}\right]\,\,
 \me^{- \frac{u}{\xi}}  \,\,\me^{ u \sqrt{ \xi^{-2}  + q^2_w} }
\label{eq:I-}
\eea{}
Defining 
\be{}
\mathcal{D}(u) = \frac{f (\rho_0 + \rho(0, 0))}{4 \pi B \rho^2_0} \me^{- u/\xi}
\label{eq:Du}
\ee{}
we can write the above expressions in compact notation
\be
I(u^{\pm}, q_w) = \mathcal{D}(u^\pm) \left[1\pm \frac{\xi^{-1}}{\sqrt{\xi^{-2}  + q^2_w}}\right] \,\me^{\mp u^\pm \sqrt{ \xi^{-2}  + q^2_w} }
\label{eq:Ipm}
\ee
where $u^{\pm}$ denotes positive and negative $u$, respectively. Note that the function in \eqref{eq:Ipm} is even in $q_w$.
We now evaluate the integral in \eqref{eq:defInvFourier}, 
\bea{}
\rho(u^\pm, w) & = & \mathcal{D}(u^\pm) \intpminf \left[1\pm \frac{\xi^{-1}}{\sqrt{\xi^{-2}  + q^2_w}}\right] \me^{\mp u^\pm \sqrt{ \xi^{-2}  + q^2_w} } \,\,\,\me^{i q_w w} \,dq_w  \nn \\
& = & 2 \mathcal{D}(u^\pm) \intpinf \left[1\pm \frac{\xi^{-1}}{\sqrt{\xi^{-2}  + q^2_w}}\right] \,\,\me^{\mp u^\pm \sqrt{ \xi^{-2}  + q^2_w} }\,\, \cos{(wq_w)} \,dq_w \nn \\
& = & 2 \mathcal{D}(u^\pm)\left[ \intpinf \me^{\mp u^\pm \sqrt{ \xi^{-2}  + q^2_w}  } \,\, \cos{(wq_w)} \,dq_w
\pm  \xi^{-1} \intpinf \frac{\me^{\mp u^\pm \sqrt{\xi^{-2}  + q^2_w}}}
{\sqrt{\xi^{-2}  + q^2_w}} 
\cos{(wq_w)} \,dq_w \right] \nn \\
\label{eq:rhouw+}
\eea{}

\noindent
From Erdelyi~\cite{Erdelyi} -- Integral Transform Table (1.4.26), \\
\bea{}
 \intpinf \me^{-u \sqrt{\xi^{-2} + q^2_w}} \,\,\cos{(wq_w)} \,dq_w \\ & = & \mathcal{I}_1(u, w) \; \text{  for  } w > 0 \nn \\
 & = &  \f{u}{\xi} \l u^2 + w^2 \r^{-\frac{1}{2}} K_1 \left[ \xi^{-1} \l u^2 + w^2 \r^{\frac{1}{2}} \right]
\label{eq:Erdelyi1.4.26}
\eea{}
From Erdelyi~\cite{Erdelyi} -- Integral Transform Table (1.4.27), \\
\bea{}
 \intpinf \frac{\me^{-u \sqrt{\xi^{-2} + q^2_w}}}{\sqrt{\xi^{-2} + q^2_w}} \,\,\cos{(wq_w)} \,dq_w & = & \mathcal{I}_2(u, w) \; \text{  for  } w > 0 \nn \\
 & = & K_0 \left[ \xi^{-1} \l u^2 + w^2 \r^{\frac{1}{2}} \right]
\label{eq:Erdelyi1.4.27}
\eea{}
where $K_0$ and $K_1$ are the modified Bessel functions of the second kind. 
\\

\noindent
Therefore, for $u > 0$,
\be{}
\rho(u, w) = 2 \mathcal{D}(u) \left[ \mathcal{I}_1(u, w) + \frac{\mathcal{I}_2(u, w)}{\xi} \right]  
\label{eq:rhofull_u+ve}
\ee{}
and for $u < 0$,
\be{}
\rho(u, w) = 2 \mathcal{D}(u) \left[ \mathcal{I}_1(u, w) -  \frac{\mathcal{I}_2(u, w)}{\xi} \right]  
\label{eq:rhofull_u-ve}
\ee{}
where,
\bea{}
\mathcal{I}_1(u, w) & = & \frac{u/\xi}{\sqrt{u^2 + w^2}} K_1 \left[ \frac{\l u^2 + w^2 \r^{\frac{1}{2}}}{\xi} \right]  \nn \\
\mathcal{I}_2(u, w) & = & K_0 \left[ \frac{\l u^2 + w^2 \r^{\frac{1}{2}}}{\xi} \right] \nn \\
\mathcal{D}(u) & = & \frac{(\rho_0 + \rho(0, 0)) f}{2 \pi v_0 \xi} \me^{- \frac{u}{\xi}}
\label{eq:I1_I2_Du}
\eea{}

\begin{figure*}[h]
\centering
\includegraphics[width=0.85\textwidth]{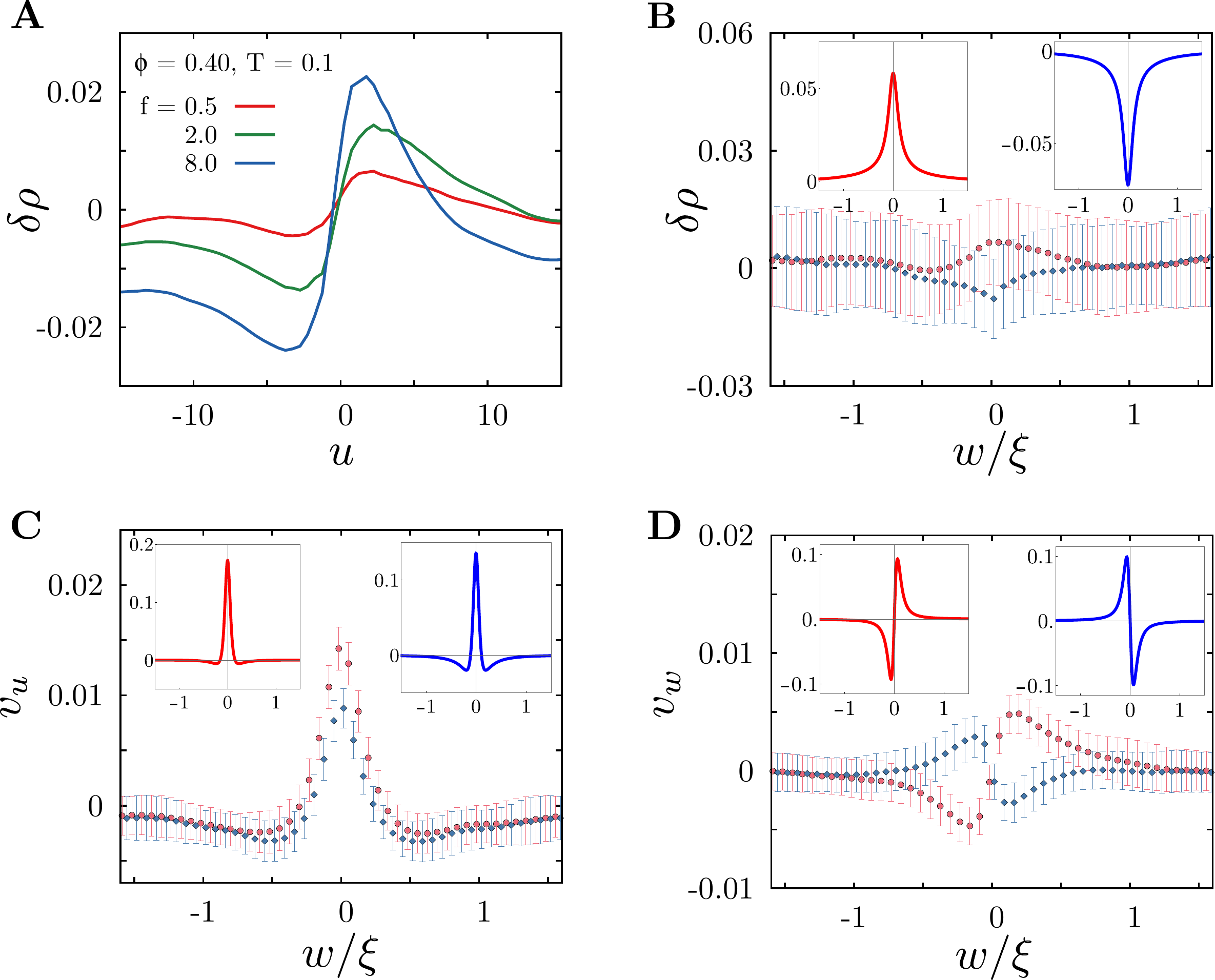}
\caption{{\bf A} Excess density profile $\delta \rho$ in the frame of the moving active particle obtained from the simulations at $w = 0$, for different active forces, $f$, and $\phi = 0.40$. The decay length scale of the density wake ($\xi$) decreases as $f$ increases. 
{\bf B}-{\bf D} The excess density, u-velocity and w-velocity profiles (for $\phi = 0.45$, $T = 0.1$) for $\frac{u}{\xi} \to 0^{+}$ (red), and $\frac{u}{\xi} \to 0^{-}$ (blue), as a function of $\frac{w}{\xi}$, with the predictions of the linear hydrodynamic theory in the inset.
}
\label{fig:SI05a}
\end{figure*}

\noindent
The fore-aft asymmetry of the density profile is most apparent when we set $w=0^+$. In this case,
for $u> 0$,
\be{}
\rho(u, 0^+) = \frac{(\rho_0 + \rho(0, 0)) f}{\pi v_0 \xi^2} \me^{- \frac{u}{\xi}} \left[ K_1 \l \frac{u}{\xi} \r + K_0 \l \frac{u}{\xi} \r \right]\,\,\,\,\,\,\,\mbox{and}
\label{eq:rho_y0+_+ve}
\ee{}
\noindent
for $u < 0$,
\be{}
\rho(u, 0^+) = - \frac{(\rho_0 + \rho(0, 0)) f}{\pi v_0 \xi^2} \me^{- \frac{u}{\xi}} \left[ K_1 \l \frac{|u|}{\xi} \r + K_0 \l \frac{|u|}{\xi} \r \right]
\label{eq:rho_y0+_-ve}
\ee{}
Using the asymptotic expansion 
\be
K_{1,0}(z) \sim \sqrt{\f {\pi}{2\vert z\vert}} e^{-z} + \ldots, \,\,\,\,\, \mbox{for large $z$}
 \ee
we see immediately that the excess density profile in 
the moving frame is fore-aft asymmetric. While the profile decays exponentially following a pile up in the region $u>0$, there is a slower power-law decay in the region $u<0$, going as ${\vert u\vert}^{-1/2}$. This is shown in Fig.\,\ref{fig:SI05a}.

In a similar way, we obtain the density profile in the direction transverse to the direction of motion,




\be{}
\rho(0^+, w) = \frac{(\rho_0 + \rho(0, 0)) f}{\pi v_0 \xi^2} K_0 \l \frac{\vert w\vert}{\xi} \r
\label{eq:rho_u0+}
\ee{}


\subsection{Two active particles moving through the compressible medium in $d=2$}
\label{sec:lineartwo}

We will now be interested in how two active particles
at ${\R}_1, {\R}_2$, moving through the compressible medium, affect each others dynamics. Let us again, take the limit of large persistence time - in this limit
the orientations $\n_1$ and $\n_2$ are time independent.
 To $\mathcal{O}(\rho)$, these particles, individually, leave a time dependent asymmetric wake described by $\rho(u,w, t)$
in \eqref{eq:rhofull_u+ve}, \eqref{eq:rhofull_u-ve}.

To compute the back reaction of this wake on the movement of the active particles, we need to include terms to 
$\mathcal{O}(\rho)$ in 
\eqref{eq:activepart}. Thus, 
\bea
\label{eq:twopart1}
 \gamma {\dot \R}_1 & = & f \n_1 -  C \rho_0 \nabla \rho \Big \vert_{\mbox{self}_1}  - C \rho_0 \nabla \rho \Big \vert_{1 \leftarrow 2} \\
\gamma {\dot \R}_2 & = & f \n_2 - C \rho_0 \nabla \rho \Big \vert_{\mbox{self}_2} - C \rho_0 \nabla \rho \Big 
\vert_{2 \leftarrow 1} 
\label{eq:twopart2}
\eea
These equations may be cast in terms of the relative coordinate $\R_{rel} = \R_1-\R_2$ and the centre of mass $\R_{cm} = (\R_1+\R_2)/2$. We look at the following scattering geometries and initial conditions -
\\

\begin{enumerate}
\item {\bf{Active particles 1 and 2 oriented along 
the fixed $\hat \x$ direction, with initial positions along the x-axis (particle 1 ahead of particle 2)}}

Note that the two active particles are identical, and so their self contributions {\it to this order} are the same.
Therefore, in terms of the relative coordinate $X_{rel} = X_1-X_2 >0$ and the centre of mass $X_{cm} = (X_1+X_2)/2$, \eqref{eq:twopart1},
\eqref{eq:twopart2} can be recast as,
\be
 \gamma {\dot X}_{rel}  =   - C \rho_0 \left( \p_x\rho \Big \vert_{1 \leftarrow 2} -
 \p_x \rho \Big 
\vert_{2 \leftarrow 1}  \right)
\label{eq:relon1}
\ee
and 
\be
\gamma {\dot X}_{cm}  =   f - C \rho_0 \p_x \rho \Big \vert_{\mbox{self}} -
\f{C \rho_0}{2} \left( \p_x\rho \Big \vert_{1 \leftarrow 2} +
 \, \p_x \rho \Big 
\vert_{2 \leftarrow 1}  \right) 
\label{eq:xcm}
\ee
We focus on the relative coordinate $X_{rel}$. The effect of particle 2 on 1 can be obtained from \eqref{eq:rho_y0+_+ve}
\be
\p_u\rho \Big \vert_{1 \leftarrow 2} = \f{(\rho_0+\rho_2)\,f}{v_0 \sqrt{2\pi}\xi^3} \left(\f{u}{\xi}\right)^{-3/2} e^{-2u/\xi} \left[1+4 \f{u}{\xi}\right]
\ee
where $u = X_1-X_2$. 
The effect of particle 1 on 2 is  obtained from \eqref{eq:rho_y0+_-ve},
\be
\p_u\rho \Big \vert_{2 \leftarrow 1} = -\f{(\rho_0+\rho_1)\,f}{v_0 \sqrt{2\pi}\xi^3} \left(\f{\vert u\vert}{\xi}\right)^{-3/2} 
\ee
where $\vert u\vert = \vert X_2-X_1\vert$.
Since the two active particles are identical, $\rho_1=\rho_2$, and so
\be
 \gamma {\dot X}_{rel}  =   - C \rho_0 \f{(\rho_0+\rho_1)\,f}{v_0 \sqrt{2\pi}\xi^3} \left(\f{X_{rel}}{\xi}\right)^{-3/2} \left[ 1 -  e^{-2\f{X_{rel}}{\xi}} \left(1+4\f{X_{rel}}{\xi}\right) \right]
\label{eq:relon2}
\ee
This shows that ${\dot X}_{rel}<0$, i.e., the inter-particle separation {\it decreases} in time, starting from an initial value $X_{rel}(0)$. The speed of approach of particle 2 towards particle 1, {\it increases} as $X_{rel} \to 0$, i.e., the trailing  particle catches up with the moving leading particle. This is a signature of {\it nonreciprocal sensing}.

We may extract the scaling behaviour from the dominant terms in 
\eqref{eq:relon2}. When $X_{rel} \gg \xi$, one may drop the 
exponential term, to obtain,
\be 
X_{rel} \propto \vert t-t_{*}\vert^{2/5}\,,
\ee
which crosses over to a late time behaviour,
\be
X_{rel} \propto \vert t-t_{*}\vert^{2/3}
\ee
on approaching the leading particle 1.

\item {\bf{Active particles 1 and 2 oriented along 
the fixed $\hat \x$ direction, with initial positions along the y-axis (particle 1 to the left of particle 2)}}

\be
 \gamma {\dot Y}_{rel}  =   - C \rho_0 \left( \p_y\rho \Big \vert_{1 \leftarrow 2} -
 \p_y \rho \Big 
\vert_{2 \leftarrow 1}  \right)
\label{eq:reloff1}
\ee
and 
\be
\gamma {\dot Y}_{cm}  =   f - C \rho_0 \p_y \rho \Big \vert_{\mbox{self}} -
\f{C \rho_0}{2} \left( \p_y\rho \Big \vert_{1 \leftarrow 2} +
 \, \p_y \rho \Big 
\vert_{2 \leftarrow 1}  \right) 
\label{eq:ycm}
\ee
The symmetry of the problem ensures that the last term in \eqref{eq:ycm} will be zero. To evaluate $Y_{rel}$, we compute
\bea
\p_y\rho \Big \vert_{1 \leftarrow 2} & = & \f{f (\rho_0 + \rho(0,0))}{2 \sqrt{2 \pi} v_0 \xi^3} \left( \f{\vert Y_{rel} \vert}{\xi}  \right)^{-3/2} e^{-\f{\vert Y_{rel} \vert}{\xi}} \left[ 1 + 2 \f{\vert Y_{rel} \vert}{\xi} \right]  
\nn \\
 & = & - \p_y\rho \Big \vert_{2 \leftarrow 1}
\eea
which, when substituted in \eqref{eq:reloff1}, gives   
\be
{\dot Y}_{rel}  =  - \f{C}{\gamma} \rho_0 \f{f^{'} (\rho_0 + \rho(0,0))}{2 \sqrt{2 \pi} v_0 \xi^3} \left(\f{\vert Y_{rel} \vert}{\xi}\right)^{-3/2} e^{-\f{2\vert Y_{rel} \vert}{\xi}}  \left[ 1+4\f{\vert Y_{rel} \vert}{\xi} \right]
\label{eq:reloff2}
\ee
Unlike in Case\,(1), particles far separated along $y$ do not feel a dynamical attractive force. However, if the y-distance between particles is of order $\xi$, then they will feel an effective attraction, which will make their trajectories {\it converge} towards each other.
At late times, when $Y_{rel} \ll \xi$, one may obtain the scaling behaviour
\be
Y_{rel} \propto \vert t-t_{*}\vert^{2/5}
\ee
which is a slower approach than Case\,(1).

\end{enumerate}

\section{Linearised hydrodynamics theory in $d=1$ : density profiles}
\label{sec:1and3}

The linearised theory shows that the excess density in $d=1$ is very different from higher dimensions, because of the absence of a back flow in the medium. We will verify this against  exact numerical
solutions of the nonlinear equations in $d=1$.

We first recall the linearised equations written in the comoving coordinate $u$ of the motile particle,
\be
- v_0 \, \pu \rho - B \,\rho^2_0 \, \pu^2 \rho + f\, \pu \l (\rho_0 + \rho) \,\delta(u) \r = 0
\label{eq:linearisedEq1}
\ee
As before, the solution is obtained by Fourier transformation,
\be{}
\fou{\rho}(q_u) = \frac{i f q_u (\rho_0 + \rho(0, 0))}{i v_0 q_u - B \rho^2_0 q_u^2}\,.
\label{eq:solnFourier1}
\ee{}
which upon inverse fourier transforming gives, 
\bea{}
\rho(u) & = & \frac{1}{2 \pi} \intpminf \fou{\rho}(q_u) \,\me^{i q_u u}  \,dq_u  \nn \\
& = & - \frac{i f (\rho_0 + \rho(0, 0))}{2 \pi B \rho^2_0} \intpminf
\frac{\me^{i q_u u}}{q_u - \frac{i v_0}{B \rho^2_0}}   \,dq_u 
\label{eq:defInvFourier1}
\eea{}
The integrand has a simple pole at
\be{}
q_{+}  =  \frac{iv_0}{B \rho^2_0} = i \xi^{-1}
\label{eq:qpm1}
\ee{}
where the length scale $\xi=B\rho_0^2/v_0$. Note that
the sgn\,$\operatorname{Im}(q_+) = +1 $.

For u $> 0$, the integrand will tend to zero only when $\operatorname{Im}(q_u) > 0$. For u $< 0$, the integrand will tend to zero only when $\operatorname{Im}(q_u) < 0$.
Thus, the contour of the integral should traverse the upper half plane around $q_+$ when u $> 0$, and in the lower half plane when u $< 0$.



This immediately shows that the excess density \\
\bea 
\label{eq:rho+}
\rho(u) & = & 
 \frac{f (\rho_0 + \rho(0, 0))}{ B \rho^2_0} \me^{- \frac{u}{\xi}} \,\,\,\,\,\, \mbox{for $u>0$} \\
 & = & 0 \,\,\,\,\,\, \mbox{for $u<0$}  \nn
\eea

Unlike the 2d density profile, there is no trailing wake in $d=1$.

\section{Numerical solution of nonlinear hydrodynamic equations in $d=1$}
\label{sec:nonlinear}

\begin{figure*}[h]
\centering
\includegraphics[width=0.85\textwidth]{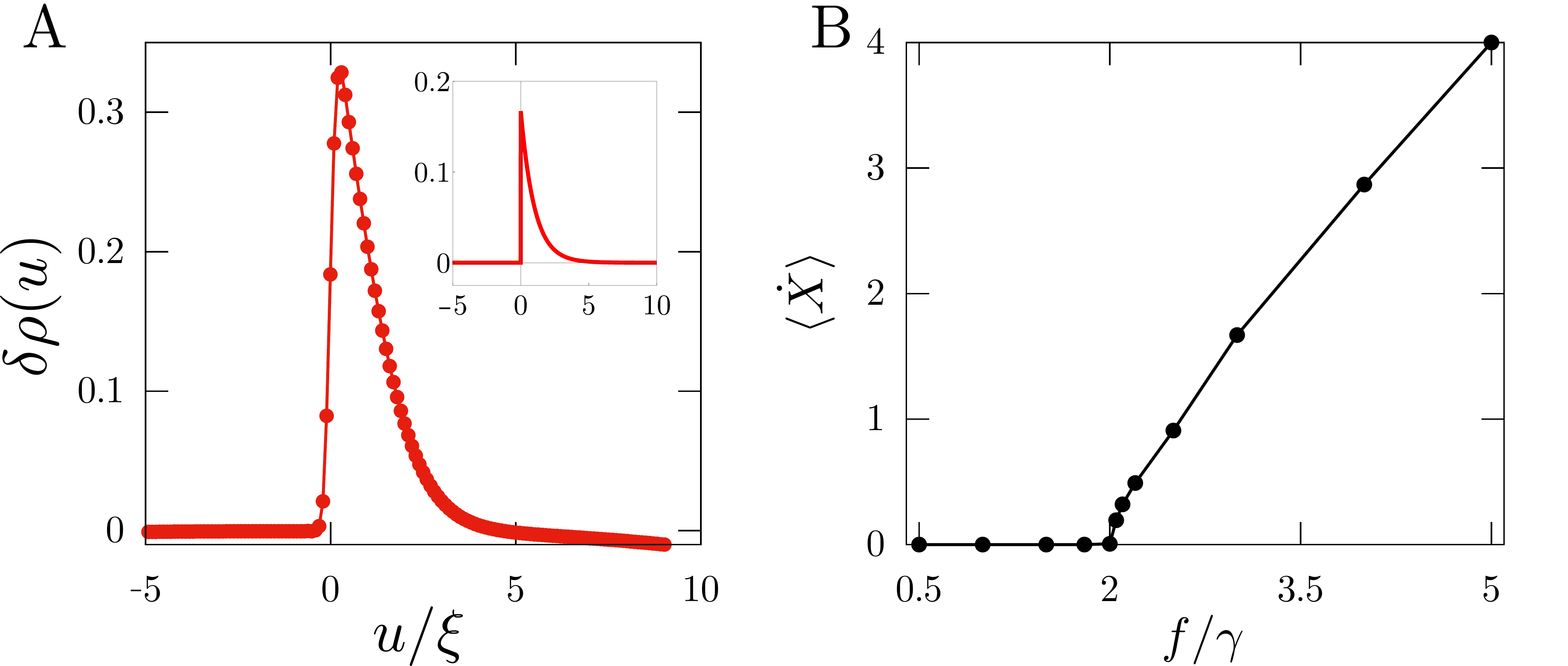}
\caption{{\bf (A)} Density profile in the co-moving frame obtained from an `exact' numerical solution of the nonlinear equation in $d=1$. Inset shows analytical result from the linear theory.  Parameter values used are $B/\Gamma=6$, $C/\gamma=10$ and $f/\gamma=2$. {\bf (B)} Dynamical transition of the active particle as a function of propulsion force $f$. This behaviour is also observed in the simulations in 2-dimensions (Fig.~4C in main text). The parameter values used are $B/\Gamma=8.6$ and $C/\gamma=25.5$. The space and time discretisation used are $dx=0.5$ and $dt=0.05$, respectively.}
\label{fig:SI06}
\end{figure*}

We numerically solve the 1-dimensional non-linear equation corresponding to Eq.\,\ref{eq:rhoevo2} and
Eq.\,\ref{eq:activepart}, with no noise.
We have used finite volume discretization with an exponential scheme~\cite{allen1955} for the convective flux as implemented in NIST-FiPy~\cite{guyer2009}. For the stability of the numerical scheme, we consider a diffusion term in the density field evolution $D \partial_x^2{\rho}$ with a very small diffusion constant ($D\sim 10^{-3}$). According to the exponential scheme, the convective problem is discretized and solved in each grid using an approximate form of the convective transport with constant coefficients $\partial_x\l\v\rho-D\partial_x{\rho}\r=\partial_x{J_x}=0$. Thus, the local (over a grid) analytical solution for the density becomes exponential. This analytical expression is then used to estimate the flux. The flux at the interface of the $i^{th}$ and $(i-1)^{th}$ grid in this scheme is given by
\bea
J_x & = & \frac{D}{\Delta x} \left( B \rho^{i-1} - A \rho^i\right) \nn\\
A & = & \frac{Pe}{e^{Pe} - 1} \nn\\
B & = & A + Pe \, ,
\eea
where $Pe=\frac{\v\Delta x}{D}$ is the P\'{e}clet number and $\Delta x$ is the grid spacing. This scheme guarantees positive solutions and has low diffusive error as the flux is formulated using the exact solution. A first order temporal discretization was used in combination with a {\it sweep} between each iteration using Newton's method. 
Numerical solution of the nonlinear equation shows perfect agreement with the linear theory for $d=1$ and that the density profile has the exponential decay in the front without any trailing wake, consistent with the linear theory (Fig.~\ref{fig:SI06}A). We also capture the dynamical transition (Fig.~\ref{fig:SI06}B) of the active particle with increasing propulsion force $f$. 

{}